%% file: main.tex
\documentclass[
    aps,
    prd,
    reprint,
    amsmath,
    amssymb,
    nofootinbib,
    longbibliography,
    preprintnumbers,
    floatfix
]{revtex4-2}
\usepackage{graphicx}
\usepackage{amsthm}
\usepackage{tikz-cd}
\usepackage{tikz-3dplot}
\usepackage{mathrsfs}
\usepackage{mathtools}
\usepackage{dsfont}
\usepackage{blkarray}
\usepackage{todonotes}
\usepackage{lmodern}
\usepackage{hyperref}
\usepackage[dvipsnames]{xcolor}
\usepackage{bm}
\usepackage[T1]{fontenc}
\usepackage{needspace}

\usepackage[most]{tcolorbox}
\newtcolorbox{movebox}{breakable, enhanced, colback=white, colframe=black, boxrule=0.4pt, arc=4pt, left=6pt, right=6pt, top=6pt, bottom=6pt, width=\columnwidth}
\usetikzlibrary{arrows.meta, decorations.pathmorphing, decorations.markings,bending,positioning,fpu}

\usepackage[caption=false]{subfig}

\begin{document}

\preprint{QMUL-PH-26-34}

\title{Direct-to-Symbol Integration from Landau Analysis}

\author{Craig Larkin}
\email{craig.larkin@ed.ac.uk}
\affiliation{%
    Higgs Centre for Theoretical Physics, 
    School of Physics and Astronomy, 
    The University of Edinburgh 
    \\ Edinburgh EH9 3FD, Scotland, UK
}

\author{Andrew J.~McLeod}
\email{andrew.mcleod@ed.ac.uk}
\affiliation{%
    Higgs Centre for Theoretical Physics, 
    School of Physics and Astronomy, 
    The University of Edinburgh 
    \\ Edinburgh EH9 3FD, Scotland, UK
}

\author{Andrzej Pokraka}
\email{a.m.pokraka@uva.nl}
\affiliation{%
        Institute of Physics, University of Amsterdam, 
        Amsterdam, 1098 XH, The Netherlands
}

\author{Lecheng Ren}
\email{lecheng.ren@qmul.ac.uk}
\affiliation{%
    Centre for Theoretical Physics, 
    Department of Physics and Astronomy, 
    \\ Queen Mary University of London, E1 4NS, UK
}

\begin{abstract}
We propose a Landau-analysis-inspired strategy for directly computing the symbol of integrals that evaluate to multiple polylogarithms in twisted cohomology. The central idea is to identify all ways in which singular hypersurfaces, twisted hypersurfaces, and integration boundaries can interact when external parameters are varied, giving rise to logarithmic or algebraic branch points. By tracking how an integral is modified when analytically continued around each of its branch points, we can recursively construct its symbol. We illustrate this approach by outlining an efficient algorithm for computing the symbol of finite integrals over twisted hyperplane arrangements, when they are expanded around special values of the twist parameter. This algorithm can be used to compute the (transcendental) integrands of cosmological correlators in conformally coupled theories to any order in the twist expansion.
\end{abstract}

\maketitle

\section{Introduction}

Integrals involving non-integer powers of rational functions appear in a variety of mathematical and physical contexts, ranging from the study of generalized hypergeometric functions to dimensionally regulated quantum field theory. These integrals admit a natural description in terms of twisted cohomology, and in recent years physicists have made increasing use of tools from intersection theory to study their properties and relations~\cite{Muhlbauer:2022ylo,Frellesvig:2019uqt,Mizera:2017rqa,Mandal:2024wun,Mizera:2019gea}. At the same time, renewed interest in Landau analysis and the use of Picard--Lefschetz theory for understanding the singularity structure of such integrals has given rise to powerful new tools~\cite{Fevola:2023kaw,Fevola:2023fzn,Caron-Huot:2024brh,Correia:2025wtb,Hannesdottir:2025bss,Chestnov:2026mpo}. These developments have provided valuable insight into the types of special functions a given integral will evaluate to, and the global analytic structure it can be expected to exhibit.

In this paper, we outline a novel strategy for directly constructing what sequences of discontinuities will appear in twisted integrals of the form
\begin{equation}
\mathcal{I}({\bf t},{\epsilon})
=
\int_{\Gamma({\bf t})}
\frac{{\prod_i T_i({\bf x},{\bf t})^{\alpha_i({\epsilon})}}}{
\prod_j D_j({\bf x},{\bf t})^{\nu_j}} \, d^n{\bf x}  
\, ,    \label{eq:intro_integrals}
\end{equation}
where the $T_i({\bf x},{\bf t})$, $D_j({\bf x},{\bf t})$, and $\alpha_i({\epsilon})$ are polynomials, $\epsilon$ is a twist parameter, the $\nu_j$ are integers, and the boldface variables ${\bf x}$ and ${\bf t}$ represent vectors whose components correspond to integration variables and external parameters. While such integrals generically evaluate to generalized hypergeometric functions, they can also evaluate to multiple polylogarithms~\cite{Chen,G91b,Goncharov:1998kja,Remiddi:1999ew,Borwein:1999js,Moch:2001zr} in special cases or upon expansion around certain values of the twist parameter \cite{%
    Arkani-Hamed:2015bza, 
    Hillman:2019wgh,
    Abreu:2019wzk,
    Brown:2019jng,
    Arkani-Hamed:2023kig,
    Arkani-Hamed:2023bsv,
    De:2023xue,
    Chowdhury:2023arc,
    He:2024olr,
    Gasparotto:2024bku,
    Baumann:2025qjx,
    Baumann:2024mvm,
    Glew:2025ypb,
    Chowdhury:2026dwm,
    Mazloumi:2025pmx,
    Paranjape:2026htn,
    Ferro:2026oph,
    Capuano:2026pgq,
    Glazer:2026whh,
    Baumann:2026atn,
    Pimentel:2026kqc,
    Fan:2025scu,
    Benincasa:2024lxe,
    Henn:2026lfz%
}.%

Landau analysis provides us with a geometric way to understand where singularities arise in integrals of the form~\eqref{eq:intro_integrals} via the interplay of the integrand and the integration contour~\cite{nakanishi1959,landau1959,Bjorken:1959fd}. As the external parameters in ${\bf t}$ are varied, the twisted hypersurfaces $T_i({\bf x}, {\bf t})=0$ and singular hypersurfaces $D_j({\bf x},{\bf t})=0$ can intersect themselves, each other, or integration boundaries in such a way that the integration contour cannot be deformed away from a singularity of the integrand. Moreover, Picard--Lefschetz theory dictates the form that discontinuities with respect to these singularities can take, when defined as the difference between the integral before and after analytic continuation around a given singular hypersurface $\lambda = 0$ in the space of external parameters:
\begin{align}
\text{Disc}_{\lambda}\left(\mathcal{I}({\bf t}, { \epsilon})\right) &=\left(\mathcal{M}_{\lambda}^{\circlearrowleft}-1\right)\mathcal{I}({\bf t},{ \epsilon}) \, .\label{eq:disc_def}
\end{align}  
Namely, the Picard--Lefschetz formula states that we can always write
\begin{align}
\text{Disc}_{\lambda}
\left(
\int_{\Gamma({\bf t})}
\omega
\right)
&=
\sum_i
\kappa_i({\epsilon})
\int_{\Gamma_i({\bf t})}
\omega \, ,
\label{eq:disc_linear_comb}
\end{align}
for some coefficients $\kappa_i({\epsilon})$, where the integration contours $\Gamma_i({\bf t})$ all have the property that they become homologically trivial in the limit ${\lambda \to 0}$~\cite{pham2011singularities,Berghoff:2022mqu}.  Thus, by studying when integration contours with this property can be constructed, one can constrain where discontinuities---and even sequences of discontinuities---have the possibility of arising, thereby probing the global analytic structure of the integrals under study~\cite{Berghoff:2022mqu,Fevola:2023kaw,Fevola:2023fzn,Britto:2023rig,Hannesdottir:2024cnn,Bargiela:2026lje}.

In dimensional regularization, much of the effort that has gone into characterizing the kinematic singularities of Feynman integrals has been motivated by the construction of systems of differential equations or bootstrap ans\"atze. In this work, we start from an alternative perspective that is more closely aligned with direct integration, wherein one iteratively massages each integration into a form matching the definition of a known class of special functions~\cite{Brown:2009ta,Bogner:2012dn,Panzer:2014caa,Bourjaily:2018aeq,Bostan_2018,Bourjaily:2021lnz}. In particular, we here propose a strategy that leverages our geometric understanding of how singularities arise in integrals of the form~\eqref{eq:intro_integrals} to construct their discontinuities. Since each of these discontinuities can themselves be expressed in terms of integrals of the same class, this strategy can be applied recursively to construct all sequences of discontinuities in these integrals. 

While this strategy can in principle be applied to any twisted integral over a rational function, we here focus on examples that evaluate to multiple polylogarithms, order by order in an expansion around some value of the twist parameter. In these cases we can use discontinuity information to construct the symbol~\cite{Goncharov:2010jf} of our integral at each order in the expansion. We first outline the general strategy, and then show how it can be applied to integrals over hyperplane arrangements, such as arise in the study of cosmological correlators~\cite{Arkani-Hamed:2017fdk,Arkani-Hamed:2023kig, Arkani-Hamed:2023bsv, De:2023xue}. We provide a Mathematica implementation of this algorithm, and illustrate its efficacy on two- and three-loop cosmological correlator diagrams that evaluate to multiple polylogarithms when expanded around de Sitter cosmologies. We conclude with some comments on how we expect this strategy to generalize to integrals such as flat-space Feynman integrals that involve singular and twisted hypersurfaces of higher polynomial degree.

\section{From Landau Singularities to Symbols of PolyLogarithms}
\label{sec:landau}

We begin by reviewing how singularities can arise in the types of twisted integrals in~\eqref{eq:intro_integrals}. For fixed values of ${\bf t}$, the polynomials in the integrand define a set of twisted hypersurfaces $T_i({\bf x},{\bf t})=0$ and singular hypersurfaces $D_j({\bf x},{\bf t})=0$. Depending on the powers $\alpha_i({\epsilon})$ and $\nu_j$, the integrand can develop zeros, poles, or branch cuts on these hypersurfaces. 
The integration contour $\Gamma({\bf t})$ may also involve boundaries, which we assume can be decomposed into a set of codimension-one hypersurfaces:
\begin{equation}
\partial\Gamma({\bf t})
=
\sum_j \Gamma_j({\bf t}),
\quad
\Gamma_j({\bf t})
\subseteq
\left\{
{\bf x}:B_j({\bf x},{\bf t})=0
\right\}.
\end{equation} 
We moreover assume that, for some initial value of the external parameters, the integral is finite and defines an analytic function in a neighborhood of that parametric point.

As we vary ${\bf t}$, singularities can arise when certain conditions are realized with respect to every integration variable. The singular/twisted hypersurfaces in the integrand can trap and pinch the integration contour, or a singular/twisted hypersurface can encounter one of the boundaries of integration. More generally, singularities can also arise from degenerations of the integration contour itself, for instance when the roots of an integration boundary collide. The first two conditions, which are those relevant to Feynman integrals, are encapsulated in the Landau equations~\cite{landau1959}. 
However, while significant advances have recently been made towards finding all solutions to these equations (see for instance~\cite{Fevola:2023fzn,Caron-Huot:2024brh,Correia:2025wtb,Chestnov:2026mpo,Matsubara-Heo:2025lrq,Fevola:2024acq}), it can still prove computationally prohibitive due to the need for algebraic blow-ups~\cite{10.1063/1.1724262,Landshoff1966,Berghoff:2022mqu,Fevola:2023fzn}. To avoid this subtlety, we take inspiration from direct integration algorithms, which proceed one integration at a time and never resort to explicit blowups. In our case, we identify and construct one discontinuity at a time, by tracking and classifying the ways the $T_i({\bf x},{\bf t})=0$, $D_j({\bf x},{\bf t})=0$, and $B_j({\bf x},{\bf t})=0$ hypersurfaces can intersect. Importantly, while nontrivial intersections must occur in every integration plane for a singularity to arise, we can often track what happens in each of these planes separately. If, for each class of intersections we encounter, we are able to determine the new contribution that one gets by analytically continuing around the corresponding singular hypersurface, we can compute the relevant discontinuities using~\eqref{eq:disc_def}. This leaves us with a sum of integrals of the same type, on which the same analysis can again be carried out.

In general, there is no guarantee that this procedure will terminate; for instance, computing discontinuities of generic hypergeometric functions can produce integrals of essentially the same level of complexity~\cite{Abreu:2018sat,Abreu:2018nzy,Abreu:2019eyg,Abreu:2019wzk,Abreu:2021vhb,Abreu:2019wzk,Brown:2019jng,McLeod:2026jpz,McLeod:2026kpo,Fu:2026dqb} (see also~\cite{Henn:2014qga}).
However, the situation simplifies considerably upon expanding around a fixed value of the twist parameter. Shifting $\epsilon$ so that this expansion is around $\epsilon=0$, we write
\begin{equation}
    \mathcal{I}({\bf t},\epsilon)
    =
    \sum_{m=0}^{\infty}
    \epsilon^m\,\mathcal{I}^{(m)}({\bf t}) \, .
    \label{eq:twist_expansion}
\end{equation}
The key difference between $\mathcal{I}({\bf t},\epsilon)$ and $\mathcal{I}^{(m)}({\bf t})$ is that, at any fixed order $m$, the number of logarithmic discontinuities---by which we mean the number of discontinuities that are proportional to $2 \pi i$---is bounded.\footnote{For more details on how we are careful to differentiate between logarithmic discontinuities and logarithmic branch points,  see Appendix~\ref{app:discontinuity_vs_branch_point_type}.} This is because---as we will see in Section~\ref{section:1dexample}---factors of $2 \pi i$ can only arise in one of two ways. Either they are produced due to residue-type discontinuities, which localize the integration contour to a simple pole and thereby reduce the number of remaining integrations, or due to twist-type discontinuities, which leave the number of integrations unchanged but introduce an additional power of the expansion parameter. Consequently, at any fixed order in the expansion, the number of twist-type discontinuities is bounded by the expansion order, while the number of residue-type discontinuities is bounded by the number of integrations. A recursive procedure for computing logarithmic discontinuities is thus guaranteed to terminate at each order in the expansion.

These sequences of logarithmic discontinuities admit a particularly nice interpretation when the coefficients $\mathcal{I}^{(m)}({\bf t})$ can be expressed as linear combinations of multiple polylogarithms. Multiple polylogarithms are defined recursively by
\begin{equation}
G_{a_1,\dots,a_n}(z)
:=
\int_0^z
G_{a_2,\dots,a_n}(t)\,
d \log(t-a_1) \, , \label{eq:mpl_def}
\end{equation}
where $\mathbf{a} = (a_1,\dots,a_n) \in\mathbb{C}^n$, $z$ is a complex variable, and $G(z)=1$. When all $n$ entries of $\mathbf{a}$ are zero,~\eqref{eq:mpl_def} is divergent, and we instead define
\begin{equation}
G_{\underbrace{\scriptstyle 0,\ldots,0}_{n}}(z)
:=
\frac{1}{n!}\log^n z \, .
\end{equation}
Multiple polylogarithms carry a natural notion of transcendental weight, corresponding to the number of indices $a_i$. Thus, for this class of functions, each logarithmic discontinuity lowers the transcendental weight by one, corresponding to the removal of one logarithmic integration.

The sequences of logarithmic discontinuities that appear within a given multiple polylogarithm can be made transparent by working with its symbol, which faithfully captures the function's branch-cut structure up to contributions proportional to higher-weight transcendental constants~\cite{Goncharov:2010jf}. Given a uniform-weight multiple polylogarithm $F$, we can construct its symbol recursively from its total differential,
\begin{equation}
dF = \sum_i F_{\ell_i} \, d\log \ell_i \, ,
\label{eq:symbol_differential}
\end{equation}
where each $F_{\ell_i}$ is a multiple polylogarithm whose transcendental weight is one lower than $F$; the symbol is then defined recursively by
\begin{equation}
    \mathcal{S}(F) = \sum_i \mathcal{S}(F_{\ell_i}) \otimes \ell_i \, ,
\label{eq:symbol_recursive}
\end{equation}
where we define $\mathcal{S}(1)=1$. Iterating this relation to weight zero produces a sum of tensor products, within which each entry---or symbol letter---corresponds to the argument of a $d\log$. When read from right to left, the entries of the symbol encode successive derivatives of a multiple polylogarithm; when read from left to right, they encode successive logarithmic discontinuities (for further details, see~\cite{Duhr:2014woa}).

Turning this construction around, if we know that $\mathcal{I}^{(m)}({\bf t})$ evaluates to a multiple polylogarithm of uniform transcendental weight $w$, we can formally write its symbol as 
\begin{equation}
\mathcal{S}\left(\mathcal{I}^{(m)}({\bf t})\right)
=
\sum_i c_i(t) \left( \ell^{(1)}_i(t) \otimes \cdots \otimes \ell^{(w)}_i(t) \right) ,
\label{eq:symbol_general}
\end{equation}
where the symbol letters $\ell^{(k)}_i(t)$ are algebraic functions of the external parameters whose vanishing loci encode the logarithmic branch points of $\mathcal{I}^{(m)}({\bf t})$. If the prefactors $c_i(t)$ and letters $\ell^{(k)}_i(t)$ are themselves free of branch points, discontinuities only act on the first (or leftmost) entry of the symbol. This means that
\begin{equation}
\begin{aligned}
&\mathrm{Disc}_\lambda\left(\mathcal{I}^{(m)}({\bf t})\right)
=
\sum_i \Big[ c_i(t)\, \mathrm{Disc}_\lambda\left(\log \ell^{(1)}_i(t)\right) \\
&\qquad \quad \times \left( \ell^{(2)}_i(t) \otimes \cdots \otimes \ell^{(w)}_i(t) \right) \Big]
+ \mathcal{O}\!\left((2\pi i)^2\right) .
\end{aligned}
\label{eq:disc_symbol}
\end{equation}
Correspondingly, if we can determine where $\mathcal{I}^{(m)}({\bf t})$ gives rise to logarithmic discontinuities by studying its geometry, we can place constraints on the letters that can appear in the first entry of its symbol. This correspondence becomes slightly more subtle when algebraic symbol letters are present, since logarithmic discontinuities can arise from either logarithmic or algebraic branch points of the integral. For simplicity, in what follows we restrict our attention to examples that do not involve algebraic branch points, before returning to this issue in Section~\ref{sec:beyond_hyperplanes}.

Combining our understanding of discontinuities from Landau analysis with their action on symbols from~\eqref{eq:disc_symbol}, we can now devise a method for constructing $\mathcal{S}\left(\mathcal{I}^{(m)}({\bf t})\right)$ from the geometry of the original integral. Suppose that $\mathcal{I}({\bf t},\epsilon)$ gives rise to a logarithmic discontinuity at a position $\lambda=0$ in the space of external parameters. Equation~\eqref{eq:disc_linear_comb} tells us that we must be able to express the corresponding discontinuity as a linear combination of integrals over modified integration contours, in which either the coefficient $\kappa_i({\epsilon})$ is proportional to $2 \pi i$, or $\Gamma_i({\bf t})$ involves a residue contour that evaluates to $2 \pi i$. Equation~\eqref{eq:disc_symbol} then tells us how to interpret the remaining integral, once this factor of \(2\pi i\) has been pulled out. Namely, it represents the symbol terms from the original integral that involved a logarithmic branch point at $\lambda = 0$ in their first entry, after those first entries have been stripped off. In the absence of algebraic roots, this implies that the letter that appeared in the first entry of these symbol terms was nothing other than $\lambda$, and we can write
\begin{equation}
\begin{aligned}
\mathcal{S}\left( \frac{1}{2\pi i} \mathrm{Disc}_\lambda \mathcal{I}({\bf t},\epsilon)\right)
&=
\sum_i \delta_{\lambda, \ell^{(1)}_i(t)} \\
&\qquad \ \ \times \mathcal{S}\!\left[ \frac{\kappa_i(\epsilon)}{2\pi i} \int_{\Gamma_i(t)} \omega \right] .
\end{aligned}
\label{eq:disc_symbol_projection}
\end{equation}
where $\omega$ represents the original integrand in $\mathcal{I}({\bf t},\epsilon)$. 

As the integrals that appear on the right side of~\eqref{eq:disc_symbol_projection} are also twisted integrals over rational functions, the next letters that appear in the symbol terms can be determined in the same way. Each time we compute another logarithmic discontinuity, we accrue a new Kronecker delta function that identifies the value of the next symbol letter, until all integrations have been evaluated as residues and we have accrued $m$ factors of $\epsilon$. Summing over all such sequences of logarithmic discontinuities, we can reconstruct $\mathcal{S}\left(\mathcal{I}^{(m)}({\bf t})\right)$. As highlighted above, if we work at a fixed order in the twist expansion~\eqref{eq:twist_expansion}, we are guaranteed this recursion terminates. 

Of course, employing this strategy to compute the symbol of integrals involving complicated geometries may prove difficult. At each step, we must identify all of the configurations of twisted and singular hypersurfaces that give rise to logarithmic discontinuities. To determine the relative coefficients $c_i({\bf t})$ that will be generated by successive monodromies or discontinuities, we must also track how the integration contour is deformed as we analytically continue around each of these singular hypersurfaces. However, even when these coefficients cannot be determined, a pared-down form of this analysis may still help constrain what sequences of discontinuities arise, and hence the set of terms that can appear in the symbol, as useful input for bootstrap approaches~\cite{Hannesdottir:2024hke}. 

Impressive progress has already been made towards being able to track how integration contours in more than one dimension are deformed by the computation of sequential discontinuities~\cite{Britto:2023rig,Britto:2025wzt}. By classifying the different geometric configurations that can be encountered in a given class of integrals, and working out the corresponding discontinuities in terms of a finite set of elementary geometric moves, we aim to formulate a set of rules that can be used to build an integral's sequential discontinuities algorithmically. In the present paper, we illustrate this method in a simple class of examples, namely for twisted integrals defined over hyperplane arrangements.

\section{Twisted Hyperplane Integrals}
\label{sec:hyperplane_integrals}

We now specialize to finite integrals over hyperplane arrangements, meaning that we take the twisted hypersurfaces, singular hypersurfaces, and integration boundaries in~\eqref{eq:intro_integrals} to be affine-linear in $\bm{x}$:
\begin{equation}
\begin{aligned}
T_i(\bm{x},{\bf t})
&=
\bm{\tau}_i({\bf t})\cdot\bm{x}
+
\tau_{i,0}({\bf t}) \, ,
\\
D_j(\bm{x},{\bf t})
&=
\bm{d}_j({\bf t})\cdot\bm{x}
+
d_{j,0}({\bf t}) \, ,
\\
B_k(\bm{x},{\bf t})
&=
\bm{b}_k({\bf t})\cdot\bm{x}
+
b_{k,0}({\bf t}) \, ,
\end{aligned}
\label{eq:hyperplane_definitions}
\end{equation}
where $\bm{\tau}_i$, $\bm{d}_j$, and $\bm{b}_k$ are $n$-dimensional vectors whose components may depend on the external parameters ${\bf t}$. For simplicity, we also restrict our attention to integrals in which all $\nu_j=1$ and all $\alpha_i(\epsilon) = \beta_i \epsilon$ for some constants $\beta_i$. Such integrals naturally appear in the context of conformally coupled scalar theories in FRW cosmologies~\cite{Arkani-Hamed:2023kig, Arkani-Hamed:2023bsv, De:2023xue}, where the expansion around $\epsilon=0$ captures the dynamics of near-de Sitter inflationary cosmologies. The geometry and cohomology of hyperplane arrangements have also long been used to organize the periods and coproduct structures of polylogarithmic functions~\cite{dupont2014combinatorialhopfalgebramotivic,dupont2018relativecohomologybiarrangements,Brown:2019jng}.

It is easy to see that the coefficients $\mathcal{I}^{(m)}({\bf t})$ in the $\epsilon$ expansion of these hyperplane integrals evaluate to multiple polylogarithms. Given a region where these integrals are finite (so that the expansion in $\epsilon$ can be interchanged with integration), expanding the twist factors in the integrand generates additional logarithms at each order in $\epsilon$, without introducing additional poles:
\begin{equation}
\begin{aligned}
T_i(\bm{x},{\bf t})^{\beta_i\epsilon}
&=
\exp\!\left[
\beta_i\epsilon
\log T_i(\bm{x},{\bf t})
\right]
\\
&=
\sum_{m=0}^{\infty}
\frac{(\beta_i\epsilon)^m}{m!}
\log^m T_i(\bm{x},{\bf t}) \, .
\end{aligned}
\label{eq:single_twist_expansion}
\end{equation}
Choosing an order of integration such as $x_1,\ldots,x_n$, the rational part can then be partial-fractioned with respect to the first integration variable, giving rise to terms of the form
\begin{equation}
\int F_r(x_1,\ldots,x_n,{\bf t})\,
\frac{R_r(x_2,\ldots,x_n,{\bf t})}{
x_1-a_r(x_2,\ldots,x_n,{\bf t})
} \, dx_1 \, ,
\label{eq:hyperplane_partial_fraction}
\end{equation}
where $F_r$ is a pure multiple polylogarithm of weight $m$, $R_r$ is a rational function, and $a_r$ is a polynomial. After expressing $F_r$ in terms of multiple polylogarithms in which $x_1$ only appears as the argument $z$ (which can generally be done by employing a fibration basis~\cite{brown2006multiplezetavaluesperiods,Anastasiou:2013srw,Panzer:2014caa}), the $x_1$ integration can be evaluated using the definition~\eqref{eq:mpl_def}.\footnote{Strictly speaking, the presence of a $d\log$ integration kernel does not by itself imply that the resulting integral is a multiple polylogarithm~\cite{Duhr:2020gdd}; however, we do not expect this subtlety to arise in the present class of integrals.} The same procedure can then be applied iteratively to all remaining integrations, allowing the full integral to be evaluated in terms of multiple polylogarithms. Note that partial fractioning in this way may split a finite integral into a sum of terms that are individually divergent; however, because the original integral is finite, these divergences must cancel in the full sum. 

Despite the simplicity of the hypersurfaces that enter these integrals, a variety of singular configurations can arise as ${\bf t}$ is varied. Twisted/singular hyperplanes can pinch the integration contour away from its boundaries, intersect one or more integration boundaries, or participate in configurations in which both things happen simultaneously. In particular, a singular configuration may arise due to a pinch in one integration-variable plane, and as an endpoint configuration in another.\footnote{Conversely, because we have restricted the integration boundaries to be affine-linear, singularities will not arise from the roots of an integration boundary becoming degenerate.}

It is worth pausing on this point. For a singularity to arise at the level of the full integral, a nontrivial intersection involving the hyperplanes $T_i(\bm{x},{\bf t})=0$, $D_j(\bm{x},{\bf t})=0$, and $B_k(\bm{x},{\bf t})=0$ must arise in every integration-variable plane simultaneously. Thus, we should be able to inspect each integration variable separately and see that it is in either a pinch or endpoint configuration. Moreover, as we analytically continue around the corresponding singular hypersurface, we can track how the integration contour is deformed separately in each integration-variable plane. Turning this logic around, we can instead start from an understanding of the singular configurations and associated monodromies that can arise with respect to each integration variable, and then determine which combinations can be realized simultaneously by analytically continuing the external parameters. In this way, the problem of identifying singularities and their associated discontinuities can be reformulated as a geometric compatibility question, rather than one that requires tracking the deformation of the full, higher-dimensional integration contour.

Although a wide variety of singular configurations can arise, considerable simplifications occur once we restrict our attention to the discontinuities that are visible at symbol level. Recall
from~\eqref{eq:disc_symbol} that each successive entry in the symbol encodes a logarithmic discontinuity that is proportional to a single factor of $2\pi i$. As such, we can ignore singular configurations whose discontinuities first contribute at $(2\pi i)^2$, because such a contribution can only appear at symbol level accompanied by a contribution at $\mathcal{O}(2\pi i)$. We will see below that this eliminates most of the singular configurations that need to be considered, leaving only two elementary geometric moves that need to be tracked.

\begin{figure}
    \centering
    \subfloat[\label{subfig:2a}]{%
        \begin{minipage}[c]{0.47\columnwidth}
            \centering
            \resizebox{\linewidth}{!}{%
                \input{tikz_figs/1dim/2a.tex}%
            }
        \end{minipage}%
    }
    \hfill
    \subfloat[\label{subfig:2b}]{%
        \begin{minipage}[c]{0.47\columnwidth}
            \centering
            \resizebox{\linewidth}{!}{%
                \input{tikz_figs/1dim/2b.tex}%
            }
        \end{minipage}%
    }
    \caption{How the integration contour is deformed when we analytically continue a simple pole around an integration endpoint. (a) The path along which the simple pole at $x=a_k$ is analytically continued. (b) The new integration contour, after the analytic continuation is carried out.}
    \label{fig:1D_example_2}
\end{figure}
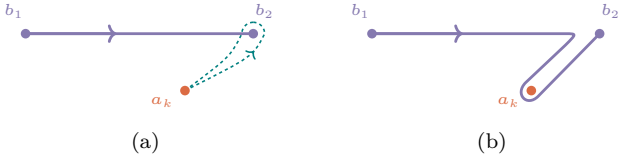

\section{Elementary Discontinuity Moves}\label{section:1dexample}

In this section we classify the geometric moves that contribute to the symbol of twisted hyperplane integrals, order by order in the expansion around $\epsilon = 0$. It turns out that nearly everything we need to understand can be discerned by studying the one-dimensional integral
\begin{equation}\label{eq:1Dexample}
        \mathcal{K}({\bf s}, {\bf a}, b_1, b_2,\epsilon)=\int_{b_1}^{b_2} dx \frac{\prod_i(x-s_i)^{\epsilon}}{\prod_j(x-a_j)}\, ,
\end{equation}
where we initially take the parameters ${\bf s} = (s_1,\dots)$, ${\bf a} = (a_1,\dots)$, $b_1$, and $b_2$ to be independent, and have set all $\beta_i=1$ for simplicity. 
By considering the different ways in which the singular and twisted hyperplanes can intersect each other and the integration boundaries, and by tracking how the integration contour is deformed as we analytically continue around the corresponding singular loci, we identify which configurations give rise to discontinuities that contribute to the symbol of (the twist expansion of) this integral. 
Moreover, by explicitly constructing the integrals that compute these discontinuities, we are able to extract sequences of symbol letters recursively (see~\cite{looprecursion} for the non-twisted case), and track the order in the expansion at which a given sequence of letters---or symbol term---will contribute. 
In subsequent sections, we will illustrate how the same moves can be uplifted to construct the symbol of hyperplane integrals involving any number of integration variables.

\vspace{.2cm}
\noindent {\bf Endpoint/Simple Pole Singularity}
\vspace{.1cm}

\noindent Let us begin by considering configurations in which a simple pole at $x=a_k$ collides with an endpoint at $x=b_2$, giving rise to a branch point at ${a_k=b_2}$ in the space of external parameters. We can compute the discontinuity with respect to this singularity by analytically continuing $a_k$ around $b_2$ counterclockwise, as shown in Figure~\ref{subfig:2a}. The contour gets deformed by this analytic continuation, as shown in Figure~\ref{subfig:2b}. To compute the discontinuity, we construct what we refer to as the difference contour by subtracting the original contour from this deformed contour. In this case, this gives us a counterclockwise-oriented residue contour centered at $x=a_k$; thus,
\begin{equation}
\begin{aligned}
\text{Disc}_{a_k-b_2}\left(\mathcal{K}\right)
&=
2\pi i\,
\underset{\ x=a_k}{\operatorname{Res}}
\left[
\frac{\prod_i(x-s_i)^{\epsilon}}
{\prod_j(x-a_j)}
\right]
\\
&=
2\pi i\,
\frac{\prod_i(a_k-s_i)^{\epsilon}}
{\prod_{j\neq k}(a_k-a_j)} \, .
\end{aligned}
\label{eq:residue-type-move}
\end{equation}
Notice that this generates an explicit factor of $2\pi i$ out front, signifying that this is indeed a logarithmic discontinuity, and that we have decreased the number of integrations by one. We refer to this type of discontinuity, in which a simple pole encircles an integration endpoint, as a \emph{residue-type move}.

\needspace{5\baselineskip}
\vspace{.2cm}
\noindent {\bf Endpoint/Branch Point Singularity}
\vspace{.1cm}

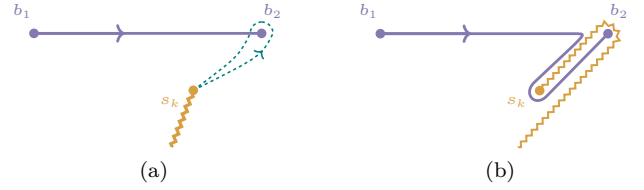
\begin{figure}
\centering
\subfloat[\label{subfig:1a}]{%
\begin{minipage}[c]{0.47\columnwidth}
\centering
\resizebox{\linewidth}{!}{%
\input{tikz_figs/1dim/1a.tex}%
}
\end{minipage}%
}
\hfill
\subfloat[\label{subfig:1b}]{%
\begin{minipage}[c]{0.47\columnwidth}
\centering
\resizebox{\linewidth}{!}{%
\input{tikz_figs/1dim/1b.tex}%
}
\end{minipage}%
}
\caption{How the integration contour is deformed when we analytically continue a twisted branch point around an integration endpoint. (a) The path along which the branch point at $x=s_1$ is analytically continued. (b) The new integration contour, after the analytic continuation is carried out.}
\label{fig:1D_example_1}
\end{figure}

\noindent We next consider situations in which a twisted branch point at $x=s_k$ collides with an endpoint at $x=b_2$, giving rise to a branch point at ${s_k=b_2}$. Similar to before, we can compute the corresponding discontinuity by continuing $s_k$ around $b_2$ counterclockwise, as shown in Figure~\ref{subfig:1a}. This time, the contour gets deformed onto a new Riemann sheet, as depicted in Figure~\ref{subfig:1b}. Crucially, this means that the portions of the integral that run from $b_2$ to $s_k$ and back differ by a prefactor $e^{2\pi i\epsilon}$, and do not cancel. Computing the discontinuity as the integral over the difference contour, we get
\begin{equation}
\begin{aligned}
&\text{Disc}_{s_k-b_2}\left(\mathcal{K}\right)
=
\left(e^{2\pi i\epsilon}-1\right)
\int_{s_k}^{b_2} \! dx\,
\frac{\prod_i(x-s_i)^{\epsilon}}
{\prod_j(x-a_j)}
\\
&\qquad \ \ \ =
2\pi i \epsilon
\int_{s_k}^{b_2} \! dx\,
\frac{\prod_i(x-s_i)^{\epsilon}}
{\prod_j(x-a_j)}
+
\mathcal{O}\!\left((2\pi i \epsilon)^2\right) \, .
\end{aligned}
\label{eq:twist-type-move}
\end{equation}
Again, the discontinuity generates a new overall factor of $2\pi i$; however, in this case the number of integrations has not decreased (although the lower endpoint has changed), and we have picked up an additional factor of $\epsilon$. Note that, since we are only interested in probing the next entry of the symbol, we discard all contributions proportional to more than one factor of $2\pi i$. We refer to this type of discontinuity, in which a twisted branch point encircles an integration endpoint, as a \emph{twist-type move}.

\vspace{.2cm}
\noindent {\bf Twisted Endpoint Singularities}
\vspace{.1cm}

\begin{figure}
    \centering
        \resizebox{0.8\linewidth}{!}{%
            \input{tikz_figs/1dim/twisted_enpoint_riemann_sheets_x1.5.tex}%
        }
    \caption{An illustration of the analytic continuation path for a simple pole around a twisted endpoint.}
    \label{fig:1D_branch_choice}
\end{figure}
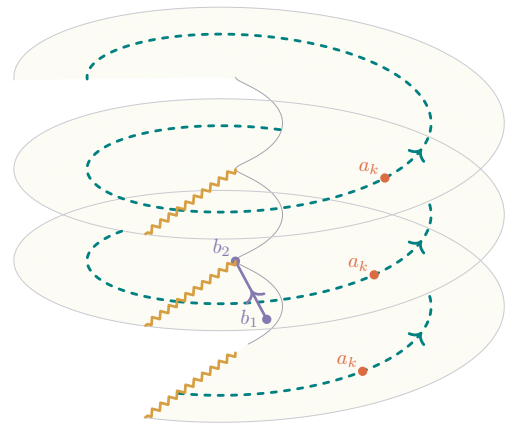

\noindent Given our analysis of twist-type moves, we generically expect to encounter integrals in which one or both of the integration endpoints sits on top of a twisted branch point. We refer to such endpoints as twisted endpoints. As it turns out, the presence of these additional twist factors does not change our previous two analyses at symbol level. To see this, we have illustrated what happens in the case of a residue-type move in Figure~\ref{fig:1D_branch_choice}. In the figure, we see that the simple pole that starts on the same Riemann sheet as the integration contour never itself encounters the contour; rather, a copy of the simple pole that started on a different Riemann sheet ends up being the one that deforms the contour. The net result is that the difference contour is identical to what one computes when the endpoint at $b_2$ is untwisted.\footnote{If we had instead placed the simple pole $a_k$ on the other side of the contour, the contour would in fact get dragged through the branch cut. In this case, the discontinuity ends up being multiplied by a factor of $e^{2\pi i \epsilon}$, but this only changes the result at higher orders in $2\pi i \epsilon$, which we are not sensitive to at symbol level.} A similar picture holds for twist-type moves around twisted endpoints. In practice, this means that we ignore whether or not an endpoint is twisted when carrying out either a residue-type or twist-type move. 

Once twisted endpoints appear, an additional discontinuity arises where the two endpoints $b_1$ and $b_2$ coincide. Consider first the case in which only one of them is twisted, say $b_1 = s_k$, while $b_2$ is not. Analytically continuing $b_1$ around $b_2$ deforms the entire contour onto the next Riemann sheet, so the discontinuity evaluates to $(e^{2\pi i \epsilon}-1)$ times the original integral. This is exactly what one obtains by applying the twist-type move~\eqref{eq:twist-type-move} with $s_k = b_1$. We can therefore treat the twisted branch point at $s_k$ as we do other twisted branch points, ignoring the fact that it coincides with the endpoint $b_1$. In other words, this situation reduces to an ordinary twist-type move.

Finally, we ought to consider the case in which both endpoints are twisted. Near the branch point at $b_1=b_2$, such an integral will take the form
\begin{equation}
\int_{b_1}^{b_2} dx\,
(x-b_1)^{\epsilon}\,
(x-b_2)^{\epsilon}\,
h(x) \, , \label{eq:double_twisted_endpoint}
\end{equation}
where $h(x)$ is single-valued in the neighborhood of the two endpoints. If we make the change of variables
\begin{equation}
x=b_1+(b_2-b_1)t \, ,
\end{equation}
the integral in~\eqref{eq:double_twisted_endpoint} becomes
\begin{equation}
(b_2-b_1)^{2\epsilon+1}
\int_0^1 dt\,
t^{\epsilon}\,
(t-1)^{\epsilon}\,
h(b_1+(b_2-b_1)t) \, .
\end{equation}
Thus, when $b_1$ and $b_2$ are analytically continued around each other, the integral picks up an overall factor of $e^{2\pi i(2\epsilon+1)} = e^{4 \pi i \epsilon}$, and the discontinuity evaluates to $e^{4\pi i\epsilon}-1 =
4\pi i\epsilon
+
\mathcal{O}\!\left((2\pi i\epsilon)^2\right)$ times the original integral. This means that the collision of two twisted endpoints can be treated as nothing more than the sum of a pair of twist-type moves---namely, of a twisted branch point at $b_1$ encircling the twisted endpoint $b_2$, and a twisted branch point at $b_2$ encircling the twisted endpoint $b_1$. As such, we do not need to introduce a new elementary move to account for this configuration either. As in the previous cases, we just need to keep track of how many twisted branch points encircle an endpoint, regardless of whether the endpoint is itself twisted. 

\vspace{.2cm}
\noindent {\bf Pinch Singularities} 
\vspace{.1cm}

\noindent Having considered the possible singularities involving integration endpoints, we now turn to pinch singularities that occur away from these endpoints. As we will see, these configurations do not contribute at symbol level. 

\begin{figure}[]
    \centering
    \subfloat[\label{subfig:3a}]{%
        \begin{minipage}[c]{0.47\columnwidth}
            \centering
            \resizebox{\linewidth}{!}{%
                \input{tikz_figs/1dim/3a.tex}%
            }
        \end{minipage}%
    }
    \hfill
    \subfloat[\label{subfig:3b}]{%
        \begin{minipage}[c]{0.47\columnwidth}
            \centering
            \resizebox{\linewidth}{!}{%
                \input{tikz_figs/1dim/3b.tex}%
            }
        \end{minipage}%
    }
\caption{How the integration contour is deformed when we analytically continue a pair of twisted branch points around each other, trapping the integration contour. (a) The analytic continuation that takes the two branch points around one another. (b) The deformed integration contour.}    \label{fig:1D_example_3}
\end{figure}
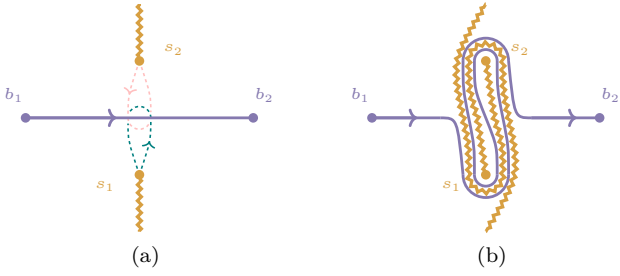

We first consider the case in which two twisted branch points pinch the integration contour. The corresponding analytic continuation is shown in Figure~\ref{subfig:3a}. In this case, the difference contour is a Pochhammer contour, and the discontinuity can be worked out to be
\begin{equation}
\text{Disc}_{s_1-s_2}\left(\mathcal{K}\right) =
\left(e^{2\pi i\epsilon}-1\right)^2\int_{s_2}^{s_1} dx\,\frac{\prod_i(x-s_i)^{\epsilon}}{\prod_j(x-a_j)} \, .\end{equation}
While this discontinuity is nonzero, we see that it first contributes at $\mathcal{O}\left((2\pi i\epsilon)^2\right)$. As there is no way to encode a discontinuity in the symbol that contributes at $\mathcal{O}\left((2\pi i\epsilon)^2\right)$ but that does not contribute at $\mathcal{O}\left(2\pi i\epsilon\right)$, such pinch singularities only encode beyond-the-symbol contributions.\footnote{Note that this observation is consistent with the fact that the symbol does not retain knowledge of the specific homotopy class of the integration contour; as such, it cannot know whether the contour gets trapped between a pair of twisted hypersurfaces.}

Similarly, we can consider pinching the integration contour between a simple pole at $x=a_k$ and a twisted branch point at $x=s_p$. The difference contour will again be a Pochhammer contour, but this time it reduces to a residue contour around $x=a_k$ evaluated on two adjacent Riemann sheets. As these two contributions differ by a factor of $e^{2\pi i\epsilon}$, we get
\begin{equation}
\begin{aligned}
\text{Disc}_{a_k-s_p}\left(\mathcal{K}\right)
&=
2\pi i
\left(e^{2\pi i\epsilon}-1\right)
\underset{x=a_k}{\operatorname{Res}}
\left[
\frac{\prod_i(x-s_i)^{\epsilon}}
{\prod_j(x-a_j)}
\right] \, .
\end{aligned}
\end{equation}
Again, we see that this discontinuity only contributes at $\mathcal{O}\left((2\pi i)^2\right)$, and thus only encodes beyond-the-symbol information.

\begin{figure}[]
    \centering
    \subfloat[\label{subfig:4a}]{%
        \begin{minipage}[c]{0.47\columnwidth}
            \centering
            \resizebox{\linewidth}{!}{%
                \input{tikz_figs/1dim/4a.tex}%
            }
        \end{minipage}%
    }
    \hfill
    \subfloat[\label{subfig:4b}]{%
        \begin{minipage}[c]{0.47\columnwidth}
            \centering
            \resizebox{\linewidth}{!}{%
                \input{tikz_figs/1dim/4b.tex}%
            }
        \end{minipage}%
    }
    \caption{How the integration contour is deformed when we analytically continue a pair of simple poles around each other, trapping the integration contour. (a) The analytic continuation that takes the two poles around one another. (b) The deformed integration contour, which differs from the original by segments that cancel pairwise, so that the difference contour integrates to zero.}
    \label{fig:1D_example_4}
\end{figure}
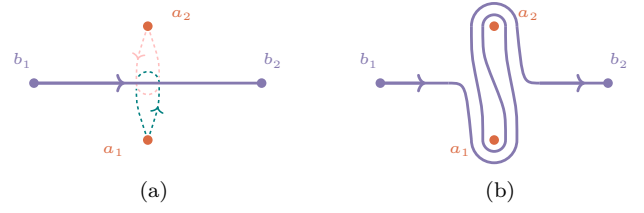

Finally, we consider the case in which two simple poles pinch the integration contour. The corresponding analytic continuation is depicted in Figure~\ref{fig:1D_example_4}. In this case, we can easily deduce that the difference contour integrates to zero, as every segment can be paired with a segment that has the opposite orientation. Thus, in contrast to the pinches involving twisted branch points, this configuration gives rise to no discontinuity at all.\footnote{Note that this is consistent with the fact that we can always partial fraction our integrand such that only a single simple pole is present in each term of the integrand.}

\vspace{.2cm}
\noindent {\bf Higher-Multiplicity Intersections} 
\vspace{.1cm}

\noindent Having covered all pairwise intersections, we finally consider configurations in which more than two singular/twisted hyperplanes intersect one of the integration boundaries. Again, these discontinuities can be understood as combinations of the simple moves we have already discussed. Suppose that $n_T$ twisted branch points and $n_D$ simple poles all intersect the endpoint at $b_2$ in the limit $\lambda \to 0$. Tracking the deformation of the contour as we analytically continue around $\lambda=0$, one can check that the resulting discontinuity is just a sum over $n_T$ twist-type moves and $n_D$ residue-type moves:
\begin{equation}\label{eq:1d-decomposed}
    \begin{split}
        \text{Disc}_{\lambda}\left( \mathcal{I} \right) &= 2\pi i \sum_{k=1}^{n_D} \left( \frac{\prod_i (a_k-s_i)^{\epsilon}}{\prod_{j\neq k}(a_k-a_j)} \right) \\
        & \! + (e^{2\pi i \epsilon}-1)\sum_{p=1}^{n_T} \left( \int_{s_p}^{b_2} dx \frac{\prod_i(x-s_i)^{\epsilon}}{\prod_j (x-a_j)} \right).
    \end{split}
\end{equation}
A similar equation holds for singularities involving the lower integration endpoint $b_1$. 

The same conclusion holds for higher-multiplicity pinch singularities, in which three or more singular/twisted hyperplanes intersect away from the integration boundaries. Any nonzero pinch of this type necessarily involves the contour being trapped between at least two singular/twisted hyperplanes, implying that the corresponding difference contour must evaluate to something proportional to at least two factors of $2\pi i$. Consequently, these contributions can only encode beyond-the-symbol information.

\vspace{.2cm}
\noindent {\bf Twisted Prefactors}
\vspace{.1cm}

\noindent Note that further logarithmic discontinuities can still be computed after the integration has been localized by a residue-type move. Referring back to~\eqref{eq:residue-type-move}, we recall that this operation leaves us with twist factors of the form $
(a_k-s_i)^\epsilon$, which introduce branch points around which we can analytically continue. Since the integration has already been localized, there is no remaining contour to deform; analytically continuing $a_k$ around $s_p$ simply multiplies the expression by $e^{2\pi i\epsilon}$. We therefore have
\begin{equation}
\mathrm{Disc}_{a_k-s_p}\!\left[\mathcal{R}_k\right]
=
\left(e^{2\pi i\epsilon}-1\right)\mathcal{R}_k \, ,
\end{equation}
where
\begin{equation}
\mathcal{R}_k
\equiv
\frac{\prod_i(a_k-s_i)^\epsilon}
{\prod_{j\neq k}(a_k-a_j)} \, .
\label{eq:prefactor_move_1d}
\end{equation}
This is simply a limiting case of a twist-type move, in which the relevant twist factor no longer depends on any integration variable. At symbol level, the discontinuity therefore contributes an overall factor of $2\pi i\epsilon$ while leaving the remaining expression unchanged. Although we could group this in with twist-type moves, it is computationally convenient to keep track of it separately; as such, we refer to this as a \emph{prefactor-type move}.

\vspace{.2cm}
\noindent {\bf Constructing the Symbol}
\vspace{.1cm}

\noindent Having identified the ways in which logarithmic discontinuities can arise, we
now put everything together to construct the symbol of $\mathcal{K}$ order
by order in the expansion around $\epsilon = 0$:
\begin{equation}
\mathcal{K}(\epsilon)
=
\sum_{m=0}^{\infty}
\epsilon^m\,\mathcal{K}^{(m)} \, .
\end{equation}
We recall from section~\ref{sec:landau} that computing a logarithmic-type discontinuity about a singular locus at $\lambda=0$ projects onto the symbol terms whose first entry is $\lambda$. To capture this, it is useful to define
\begin{equation}
C_{\lambda}\left(\mathcal{K}\right)
\equiv
\left[
\frac{1}{2\pi i}
\text{Disc}_{\lambda}\left(\mathcal{K}\right)
\right]_{\mathcal{O}\left((2\pi i)^0\right)} \, ,
\end{equation}
where the notation on the right side indicates that we retain just the part of $\text{Disc}_{\lambda}\left(\mathcal{K}\right)$ that is proportional to a single factor of $2\pi i$. Schematically, the symbol can then be recursively constructed as
\begin{equation}
\mathcal{S}\left(\mathcal{K}^{(m)}\right)
=
\left[\sum_{\lambda}
\lambda
\otimes
\mathcal{S}\left(
C_{\lambda}\mathcal{K}
\right) \right]_{\mathcal{O}\left(\epsilon^m \right)}\, ,
\label{eq:1d_symbol_recursion}
\end{equation}
where the sum runs over all possible residue-, twist-, and prefactor-type moves in $\mathcal{K}$.\footnote{In accordance with our discussion of twisted endpoints above, this sum includes twist-type moves in which the twisted branch point being continued is itself one of the integration endpoints, and both residue- and twist-type moves are counted regardless of whether the endpoint being encircled is twisted. When both endpoints are twisted, each encircling the other counts as a separate move.}

By counting the number of moves of each type in a given sequence, we can determine at the outset whether it contributes at $\mathcal{O}(\epsilon^m)$. Since $\mathcal{K}$
starts life as a one-fold integral, every complete sequence must contain
exactly one residue-type move in order to localize the integration,
together with $m$ moves of twist- or prefactor-type in order to generate
the correct power of $\epsilon$. Since each move takes us one entry further into the symbol,
this means that $\mathcal{K}^{(m)}$ has transcendental weight $m+1$.

\section{Two-Dimensional Example}\label{section:2dexample}

In the last section, we showed that the symbol of a one-fold twisted hyperplane integral can be constructed order by order in the twist expansion just by tracking twist- and residue-type moves.
We now generalize this statement to higher dimensions, and illustrate how these elementary moves work in a two-dimensional example. The full algorithm that we use to construct the symbol of $n$-fold twisted hyperplane integrals (order by order in the twist expansion) will then be presented in Section~\ref{section:algorithm}.

Recall from Section~\ref{sec:landau} that, for a singularity of an $n$-fold integral to arise, a nontrivial intersection must occur simultaneously in every integration-variable plane. Thus, for a pinch singularity to arise, there must exist at least one integration-variable plane in which the contour is trapped between two or more twisted/singular hypersurfaces. The difference contour in that plane is therefore of Pochhammer type, and can be analyzed as in the last section. As we saw there, these contributions always start at $\mathcal{O}\left((2 \pi i)^2 \right)$ (or vanish). Thus, we can continue to ignore pinch singularities when constructing the symbols of higher-dimensional twisted hyperplane integrals.  Moreover, it remains the case that higher-multiplicity intersections can be understood as sums of twist- and residue-type moves.

\begin{figure*}[htbp]
    \centering
    \subfloat[\label{subfig:left}]{%
        \begin{minipage}[b]{0.32\textwidth}
            \centering
            \input{tikz_figs/2dim/2a.tex}
        \end{minipage}%
    }
    \hfill
    \subfloat[\label{subfig:middle}]{%
        \begin{minipage}[b]{0.32\textwidth}
            \centering
            \input{tikz_figs/2dim/2c.tex}
        \end{minipage}%
    }
    \hfill
    \subfloat[\label{subfig:right}]{%
        \begin{minipage}[b]{0.32\textwidth}
            \centering
            \input{tikz_figs/2dim/2b.tex}
        \end{minipage}%
    }

    \caption{The hyperplane arrangement is shown with (a) the original integration contour, (b) the contour resulting from the discontinuity involving $D_1$, and (c) the contour resulting from the discontinuity involving $T_1$.}
    \label{fig:2D_example}
\end{figure*}
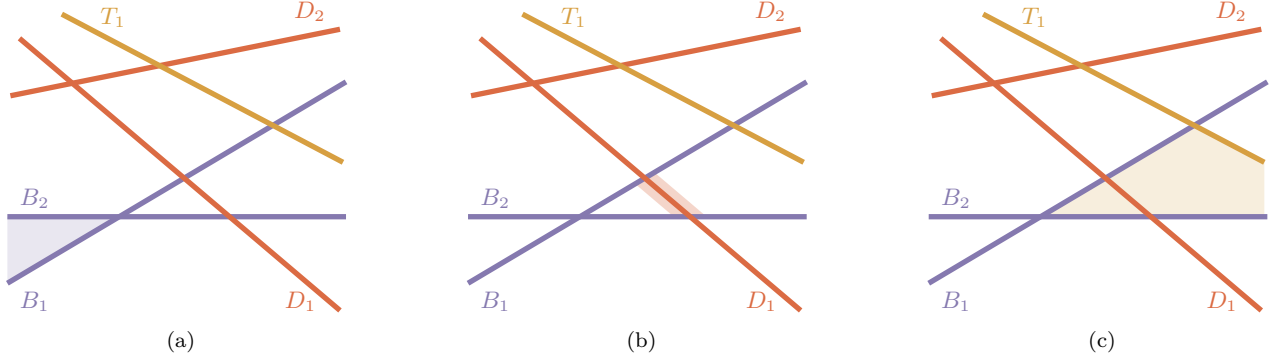

This leaves us needing to consider only discontinuities that result from twisted and singular hyperplanes intersecting with maximum-codimension integration boundaries. To reduce the number of cases we need to work out, we first map each maximum-codimension endpoint to an upper endpoint using the identity
\begin{equation}\label{eq:decomposition}
    \begin{split}
        \int_{B_1^-}^{B_1^+}\int_{B_2^-}^{B_2^+} \omega &= \int_{R}^{B_1^+}\int_{R}^{B_2^+} \omega - \int_{R}^{B_1^+}\int_{R}^{B_2^-} \omega \\
        & \qquad - \int_{R}^{B_1^-}\int_{R}^{B_2^+} \omega + \int_{R}^{B_1^-}\int_{R}^{B_2^-} \omega,
    \end{split}
\end{equation}
where $R$ denotes an arbitrary reference endpoint. The relative signs associated with different endpoints are automatically taken care of by decomposition. 

Let us now specialize to just one of the terms that will appear after this decomposition, for instance
\begin{equation}
\label{eq:2D_hyperplanes}
\int_{R}^{B_2}
\int_{R}^{B_1}
dx_1\,dx_2\,
\frac{
T_1(\bm{x})^{\epsilon}
}{
D_1(\bm{x})D_2(\bm{x})
} 
\end{equation}
where
\begin{gather*}
T_1=\bm{\tau}_1\cdot\bm{x}+\tau_{1,0} \, ,
\\
D_1=\bm{d}_1\cdot\bm{x}+d_{1,0} \, ,
\qquad
D_2=\bm{d}_2\cdot\bm{x}+d_{2,0} \, ,
\\
B_1=\bm{b}_1\cdot\bm{x}+b_{1,0} \, ,
\qquad
B_2=b_{2,2}x_2+b_{2,0} \, ,
\end{gather*}
and
$\bm{\tau}_1=(\tau_{1,1},\tau_{1,2})$,
$\bm{d}_i=(d_{i,1},d_{i,2})$, and
$\bm{b}_1=(b_{1,1},b_{1,2})$.
Note that the boundary $B_2$ does not depend on $x_1$.
An example of what the geometry of this integral might look like is depicted in Figure~\ref{subfig:left}. The upper endpoint in this integral is given by
$$
\left\{x_1,x_2\right\}\big|_{B_1 \cap B_2}
=
\left\{
\frac{
b_{1,2} b_{2,0}
-
b_{1,0} b_{2,2}
}{
b_{1,1} b_{2,2}
},
-\frac{b_{2,0}}{b_{2,2}}
\right\}.
$$
We refer to this as the relevant vertex, and now study what happens when we analytically continue one of the twisted/singular hypersurfaces around this point.

Let us first consider the case in which the simple pole hypersurface $D_1=0$ collides with the relevant vertex. In the space of external parameters, this occurs on the locus
\begin{equation} \label{eq:singular_hyperplane_boundary_intersection}
\begin{aligned}
\left.D_1\right|_{B_1\cap B_2}
&=
\frac{
d_{1,1}\left(b_{1,2}b_{2,0}-b_{1,0}b_{2,2}\right)
}{
b_{1,1}b_{2,2}
}
\\
&\qquad \quad
-\frac{d_{1,2}b_{2,0}}{b_{2,2}}
+d_{1,0}
=
0 \, .
\end{aligned}
\end{equation}
As shown in Figure~\ref{subfig:middle}, analytically continuing \(D_1=0\) around this intersection point gives rise to a difference contour that encircles ${D_1=0}$ between ${B_1=0}$ and ${B_2=0}$. The corresponding discontinuity can be written as 
\begin{align}
&2\pi i
\int_{D_1\cap B_1}^{B_2}
dx_2\,
\underset{D_1=0}{\operatorname{Res}}
\left[
\frac{
T_1(\bm{x})^{\epsilon}
}{
D_1(\bm{x})D_2(\bm{x})
}
\right]
\nonumber\\
&\qquad =
\frac{2\pi i}{d_{1,1}}
\int_{D_1\cap B_1}^{B_2}
dx_2\,
\frac{
\left(
\left.T_1(\bm{x})\right|_{D_1=0}
\right)^{\epsilon}
}{
\left.D_2(\bm{x})\right|_{D_1=0}
} \, .
\label{eq:residue-type}
\end{align}
Like in one dimension, we see that this discontinuity generates an overall factor of $2\pi i$, while localizing one integration variable. This is the higher-dimensional version of a residue-type move.

We next consider the case in which the twisted hypersurface $T_1=0$ collides with the relevant vertex. This happens where
\begin{equation} \label{eq:twisted_hyperplane_boundary_intersection}
\begin{aligned}
\left.T_1\right|_{B_1\cap B_2}
&=
\frac{
\tau_{1,1}\left(b_{1,2}b_{2,0}-b_{1,0}b_{2,2}\right)
}{
b_{1,1}b_{2,2}
}
\\
&\qquad\quad
-\frac{\tau_{1,2}b_{2,0}}{b_{2,2}}
+\tau_{1,0}
=
0 \, .
\end{aligned}
\end{equation}
As shown in Figure~\ref{subfig:right}, analytically continuing ${T_1=0}$ around this intersection point gives rise to a difference contour that is still bounded by ${B_1=0}$ and ${B_2=0}$, but now also by ${T_1=0}$. The corresponding discontinuity can be written as
\begin{equation}
\left(e^{2\pi i\epsilon}-1\right)
\int_{T_1\cap B_1}^{B_2}
dx_2\,
\int_{T_1}^{B_1}
dx_1\,
\frac{
T_1(\bm{x})^{\epsilon}
}{
D_1(\bm{x})D_2(\bm{x})
} \, .
\end{equation}
Like in one dimension, we see that this discontinuity generates an overall factor of $2\pi i\epsilon$, while leaving the number of integration variables unchanged. This is the higher-dimensional version of a twist-type move.

Together, these examples illustrate how the elementary moves identified in one dimension uplift to higher-dimensional hyperplane integrals. Like in one dimension, both types of moves generate a factor of $2\pi i$, but only the twist-type move contributes an additional factor of $\epsilon$, and only the residue-type move reduces the number of integrations by one. In both cases, the associated symbol letter is obtained by evaluating the twisted/singular hypersurface at the relevant vertex, as in~\eqref{eq:singular_hyperplane_boundary_intersection} and~\eqref{eq:twisted_hyperplane_boundary_intersection}; note that the correct symbol letter is given by the full rational function obtained from this evaluation, not just the numerator.

\section{Algorithm for Twisted Hyperplane Integrals}\label{section:algorithm}
In this section we explain the algorithm that is implemented in the Mathematica notebook attached to this paper, which can be used to efficiently compute the symbol of twisted hyperplane integrals order by order in an expansion around $\epsilon=0$. We first describe how each elementary discontinuity move is implemented, and then explain how these moves can be combined to recursively construct the symbol order by order in \(\epsilon\).

The first thing we do is fix an order of integration, and explicitly parameterize our integration contour according to this ordering. Without loss of generality, we can choose this order to  be $\{x_1,x_2,\dots,x_n\}$. We then split each $n$-fold integral into a sum of $2^n$ integrals in which the lower integration endpoint has been replaced by an arbitrary reference value $R$, using the $n$-dimensional analog of~\eqref{eq:decomposition}. At this point, each of our integrals contains just a single relevant vertex, corresponding to the intersection of all its upper endpoints. 

We now proceed by identifying all ways in which we can compute a single logarithmic discontinuity. First, we enumerate all singular hypersurfaces $D_i=0$ that can be analytically continued around the relevant vertex, and carry out a residue-type move for each of these hypersurfaces. The integral over the difference contour that one arrives at can be constructed via the following procedure:
\begin{movebox}
\textbf{Residue-Type Move}

\vspace{.1cm}
Let $x_k$ be the first integration variable on which $D_i$ depends.

\vspace{.13cm}
\begin{enumerate}
\setlength{\itemsep}{-1.8pt}
\item Replace the lower integration endpoint for $x_k$ by $D_i=0$.
\item Replace the lower endpoint of each subsequent integration variable by the intersection of the upper and lower endpoints of the immediately preceding variable. This produces the following sequence of lower integration boundaries:
{
\setlength{\abovedisplayskip}{3.4pt}
\setlength{\belowdisplayskip}{0.2pt}
\[
D_i
\longrightarrow
D_i\cap B_k
\longrightarrow
D_i\cap B_k\cap B_{k+1}
\longrightarrow
\cdots
\]}

If an intersection does not depend on the next integration variable, leave that variable's boundaries unchanged and proceed to the next variable on which it does depend.
\item Compute the residue around $D_i=0$ by evaluating the integral over $x_k$, and multiply the result by $2 \pi i$. This leaves an $(n-1)$-fold integral over an integration contour that is restricted to $D_i=0$.
\item Record $\left.D_i\right|_{B_1\cap\cdots\cap B_n}$ as the next letter in the current symbol term.
\end{enumerate}
\end{movebox}
For example, if we start with a three-fold integral and $D_i$ depends on all three integration variables, we arrive at the integral
\begin{equation}
2\pi i \int_{D_i\cap B_1\cap B_2}^{B_3} dx_3
\int_{D_i\cap B_1}^{B_2} dx_2
\, \underset{D_i=0}{\operatorname{Res}}
\left[ \, \cdots \right]\, ,
\end{equation}
where the residue of the original integrand is computed by evaluating the integral over $x_1$. This new integral captures those symbol terms of the original integral whose first entry is $\lambda=\left.D_i\right|_{B_1\cap\cdots\cap B_n}$, namely the locus where $D_i=0$ intersects the relevant vertex, after we have stripped off this first entry.

We next enumerate all twisted hypersurfaces $T_i=0$ that can be analytically continued around the relevant vertex, and carry out a twist-type move for each of these hypersurfaces. The resulting integral over the difference contour can be constructed in essentially the same way as for a residue-type move, except that we do not compute a residue that restricts the new contour to $T_i=0$:
\begin{movebox}
\textbf{Twist-Type Move}

\vspace{.1cm}
Let $x_k$ be the first integration variable on which $T_i$ depends.

\vspace{.13cm}
\begin{enumerate}
\setlength{\itemsep}{-1.8pt}
\item Replace the lower integration endpoint for $x_k$ by $T_i=0$.
\item Replace the lower endpoint of each subsequent integration variable by the intersection of the upper and lower endpoints of the immediately preceding variable. This produces the sequence
{
\setlength{\abovedisplayskip}{3.4pt}
\setlength{\belowdisplayskip}{0.2pt}
$$
T_i
\longrightarrow
T_i\cap B_k
\longrightarrow
T_i\cap B_k\cap B_{k+1}
\longrightarrow
\cdots
$$

}

If an intersection does not depend on the next integration variable, leave that variable's boundaries unchanged and proceed to the next variable on which it does depend.
\item Multiply the resulting $n$-fold integral by $e^{2\pi i\epsilon}-1$. 
\item Record $\left.T_i\right|_{B_1\cap\cdots\cap B_n}$ as the next letter in the current symbol term.
\end{enumerate}
\end{movebox}
For example, if we start with a three-fold integral and $T_i$ depends on all three integration variables, the difference contour is schematically parameterized by
\begin{equation}
\left(e^{2\pi i\epsilon}-1\right)
\int_{T_i\cap B_1\cap B_2}^{B_3} \!\!\! dx_3
\int_{T_i\cap B_1}^{B_2} \! dx_2
\int_{T_i}^{B_1} dx_1
\left[\,\cdots \right] \, .
\end{equation}
At symbol level, $e^{2\pi i\epsilon}-1=2\pi i\epsilon+\mathcal{O}\left((2\pi i\epsilon)^2\right)$, so each twist-type move contributes a factor of $2\pi i \epsilon$ while leaving the number of integrations unchanged. The remaining integral captures those symbol terms of the original integral whose first entry is $\lambda=\left.T_i\right|_{B_1\cap\cdots\cap B_n}$, namely the locus where $T_i=0$ intersects the relevant vertex, after we have stripped off the first entry. The additional factor of $\epsilon$ also increments the order in the twist expansion at which these contributions will appear. 

Finally, we identify all twisted prefactors $P_i^{\epsilon}$ that no longer depend on any integration variables. Such factors naturally arise after residue-type moves remove their dependence on integration variables. Since these factors do not depend on the integration variables, we can compute logarithmic discontinuities with respect to them without modifying the remaining integral:
\begin{movebox}
\textbf{Prefactor-Type Move}

\vspace{.1cm}
Let $P_i^{\epsilon}$ be a twisted prefactor that no longer depends on any integration variables.

\vspace{.13cm}
\begin{enumerate}
\setlength{\itemsep}{-1.8pt}
\item Identify the singular locus $P_i=0$.
\item Leave the integration contour unchanged.
\item Multiply the integral by $e^{2\pi i\epsilon}-1$.
\item Record $P_i$ as the next letter in the current symbol term.
\end{enumerate}
\end{movebox}
Since $e^{2\pi i\epsilon}-1=2\pi i\epsilon+\mathcal{O}\left((2\pi i\epsilon)^2\right)$ each prefactor-type move contributes a factor of $2 \pi i \epsilon$, while leaving the integration contour and the number of integrations unchanged. 

Having fully specified each of these moves in $n$ dimensions, we can now construct the symbol of the types of twisted hyperplane integrals described in section~\ref{sec:hyperplane_integrals}, at any order in the $\epsilon = 0$ expansion. In particular, to generate the $\mathcal{O}(\epsilon^m)$ contribution to an $n$-fold integral of this type, we need only consider ordered sequences of moves that involve exactly $n$ residue-type moves and $m$ twist- or prefactor-type moves, as this is the number of residue-type moves required to evaluate all integrations and the number of twist- and prefactor-type moves required to generate the required power of $\epsilon$. In full detail:
\begin{movebox}
\textbf{Recursive Algorithm}

\vspace{.1cm}
To generate the weight-$(n{+}m)$ symbol of an $n$-fold twisted hyperplane integral at $\mathcal{O}(\epsilon^m)$---which is the complete symbol whenever $\mathcal{I}^{(m)}({\bf t})$ has uniform weight---we must carry out all possible sequences of $n$ residue-type moves and $m$ twist/prefactor-type moves.

\vspace{.13cm}
\begin{enumerate}
\setlength{\itemsep}{-1.8pt}

\item Decompose the initial integral into a linear combination of integrals in which the lower integration endpoints have been replaced by an arbitrary reference value, and initialize each decomposed integral with an empty symbol word.

\item Generate all ways of ordering $n$ residue-type moves and $m$ twist/prefactor-type moves.

\item For the next move in a given ordering, enumerate and enact all allowed moves of the prescribed type, and append the associated letter to a given integral's symbol word.

\item Decompose each newly generated integral again, such that its lower integration endpoints have been replaced by the reference value and it has a single relevant vertex.

\item Repeat the last two steps until all $n+m$ moves have been applied.

\item Strip off the overall factor of $(2\pi i)^{n+m}\epsilon^m$, treating each factor of $e^{2\pi i\epsilon}-1$ as $2\pi i\epsilon$ in accordance with~\eqref{eq:1d_symbol_recursion}, and evaluate the remaining expression at $\epsilon=0$. Multiply the result by the associated symbol word.

\item Sum over all terms.
\end{enumerate}
\end{movebox}
We highlight that, just like in one and two dimensions, higher-multiplicity intersections require no special treatment here; the contributions that arise from these intersections are computed by summing over the individual twist- and residue-type moves that arise from each participating twisted/singular hypersurface.

Because of how we have formulated the $n$-dimensional residue- and twist-type moves, this algorithm can also be applied to integrals over non-generic hyperplane arrangements, not just generic configurations. In non-generic configurations, the twisted/singular hypersurfaces and their intersections with the boundary hypersurfaces may not depend on every integration variable; however, as described above, the unaffected integrations are simply left unchanged. Even so, these configurations may lead to terms in the recursion that fail to admit $n$ residues. For instance, if two singular hyperplanes are parallel, then computing a residue with respect to one of them causes the other to lose all dependence on the remaining integration variables, meaning that fewer singular hyperplanes remain available for subsequent residues than expected. Terms in which one cannot localize all integrations via residues encode contributions whose transcendental weight is lower than $n+m$, and correspondingly do not contribute to the maximal-weight symbol at $\mathcal{O}(\epsilon^m)$. Our algorithm discards them automatically.

\begin{figure*}[t]
    \centering

    \subfloat[\label{cc:left}]{%
        \begin{minipage}[b]{0.5\textwidth}
            \centering
            \resizebox{0.9\linewidth}{!}{%
                \input{tikz_figs/sunrise_diagram.tex}%
            }
        \end{minipage}%
    }
    \hfill
    \subfloat[\label{cc:right}]{%
        \begin{minipage}[b]{0.5\textwidth}
            \centering
            \resizebox{0.9\linewidth}{!}{%
                \input{tikz_figs/acnode_diagram.tex}%
            }
        \end{minipage}%
    }
\caption{Representative cosmological correlator diagrams used to test the recursive symbol algorithm: (a) the three-loop sunrise graph and (b) the acnode graph.}
\label{fig:cosmological_correlators}
\end{figure*}
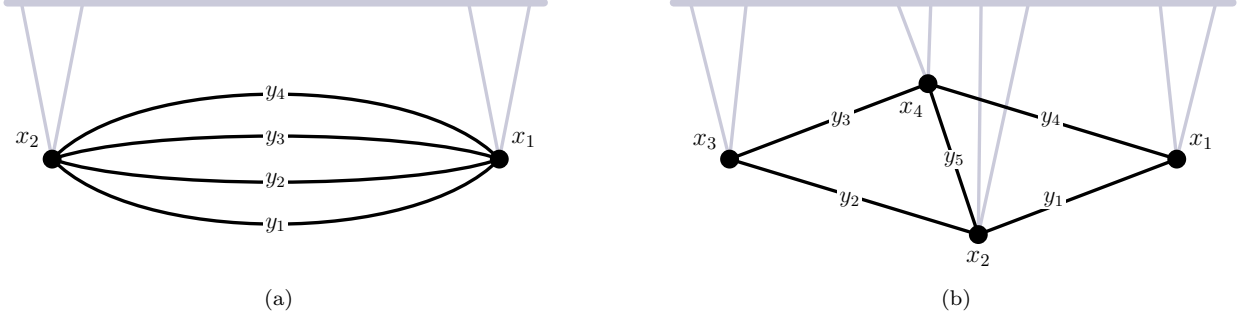

We have tested this algorithm against a variety of examples, including integrals that arise in the study of cosmological correlators~\cite{Arkani-Hamed:2023kig, Arkani-Hamed:2023bsv, De:2023xue}.\footnote{When computing cosmological correlators at symbol level, we first decompose the integral and then drop any terms whose relevant vertex is not a finite point. Such terms only generate divergent letters which must cancel among themselves, as we know the final result to be finite.} For instance, the three-loop sunrise integral depicted in Figure~\ref{cc:left} can be written as a linear combination of $24$ two-fold twisted hyperplane integrals of the type introduced in Section~\ref{sec:hyperplane_integrals}. Using the accompanying Mathematica code, we have computed the symbol of this sum of integrals through $\mathcal{O}(\epsilon^{11})$, where it has transcendental weight $13$ and contains $376{,}740$ terms. We have checked these results against HyperInt~\cite{Panzer:2014caa} through $\mathcal{O}(\epsilon^{4})$. As a second example, the acnode graph depicted in Figure~\ref{cc:right} can be written as a linear combination of 108 finite twisted hyperplane integrals. We have computed this diagram through $\mathcal{O}(\epsilon^2)$, where its symbol has transcendental weight $6$ and contains $1{,}289{,}600$ terms. These results were checked against HyperInt at $\mathcal{O}(\epsilon^0)$ and for individual twisted hyperplane contributions at $\mathcal{O}(\epsilon^1)$. For one of these 108 acnode contributions (chosen at random), computing the $\mathcal{O}(\epsilon^1)$ symbol using our algorithm took 25 seconds, whereas evaluating the same contribution in terms of multiple polylogarithms with HyperInt took 1 hour and 49 minutes on the same laptop.\footnote{To carry out this comparison, we expanded the integrand to $\mathcal{O}(\epsilon^1)$ before integrating, so that the integrand given to HyperInt involved a sum of logarithms. The time required to subsequently extract the symbol from the resulting multiple polylogarithms is not included.} Although this code is intended only as a proof of concept, it is thus already efficient enough that each of the computations described here can be carried out on a laptop.

\section{Beyond Twisted Hyperplanes}
\label{sec:beyond_hyperplanes}

Having solved the hyperplane case at symbol level, it is natural to ask how much of
this algorithm generalizes to more complicated integrals. The next class of examples we might consider is flat-space Feynman integrals in the Baikov representation, which have a singular hypersurface associated with
each propagator and a single, higher-degree twisted hypersurface (the Baikov
polynomial). However, even when these integrals evaluate to multiple
polylogarithms, extending the algorithm of Section~\ref{section:algorithm} presents two distinct challenges: correctly interpreting what a given logarithmic discontinuity
tells us about the symbol~\cite{looprecursion}, and identifying/computing all such
discontinuities in the first place. 

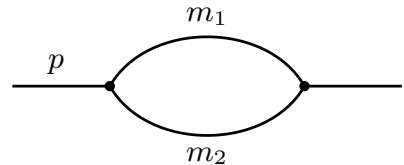
\begin{figure}[b]
    \centering
    \resizebox{0.6\linewidth}{!}{
    \begin{tikzpicture}[
        line width=0.8pt,
        baseline=(current bounding box.center)
    ]
        \coordinate (L) at (-1.0,0);
        \coordinate (R) at (1.0,0);

        \draw (-2.0,0) -- (L);
        \draw (R) -- (2.0,0);

        \draw (L)
            to[out=60,in=120]
            node[midway,above] {$m_1$}
            (R);
        \draw (L)
            to[out=-60,in=-120]
            node[midway,below] {$m_2$}
            (R);

        \fill (L) circle (1.5pt);
        \fill (R) circle (1.5pt);

        \node[above] at (-1.55,0) {$p$};
    \end{tikzpicture}}

    \caption{The massive bubble integral with internal masses $m_1$ and $m_2$.}
    \label{fig:bubble}
\end{figure}

To illustrate the first challenge, let us consider the massive bubble
Feynman integral, shown in Figure~\ref{fig:bubble}. In two spacetime
dimensions, this integral evaluates to
\begin{equation}
\begin{aligned}
\mathcal{I}_{\rm bub}^{D=2}
&=
\frac{2\pi i}{
\sqrt{s-r_+^2}\sqrt{s-r_-^2}
}
\\[-2pt]
&\qquad \quad \times
\log\!\left(
\frac{
\sqrt{s-r_+^2}-\sqrt{s-r_-^2}
}{
\sqrt{s-r_+^2}+\sqrt{s-r_-^2}
}
\right) \, ,
\end{aligned}
\label{eq:massive_bubble_2d}
\end{equation}
where $r_\pm=m_1\pm m_2$. Notice that there are algebraic branch points at
$s=r_\pm^2$, where the square roots vanish, but no logarithmic branch points at these locations. Nevertheless, computing the
discontinuity around the threshold $s = r_+^2$ via~\eqref{eq:disc_def},
one finds~\cite{Hannesdottir:2022xki}
\begin{equation}
\text{Disc}_{s-r_+^2}\!\left(\mathcal{I}_{\rm bub}^{D=2}\right)
\propto
\frac{(2\pi i)^2}{\sqrt{s-r_+^2}\sqrt{s-r_-^2}} \, ,
\label{eq:bubble_disc}
\end{equation}
which constitutes a logarithmic
discontinuity, insofar as this operation generates a new factor $2\pi i$. Thus, the existence of a logarithmic discontinuity does not by itself imply that the underlying branch point is logarithmic. Distinguishing logarithmic from algebraic branch points requires additional monodromy information, which can be obtained by computing longer sequences of discontinuities, as discussed in Appendix~\ref{app:discontinuity_vs_branch_point_type}.

The second challenge concerns the geometry itself. In the hyperplane case, every hypersurface has exactly one root in each integration-variable plane, and its position depends linearly on the remaining integration variables. Hypersurfaces of higher degree instead have several roots in each plane, whose positions depend algebraically on the remaining variables. As the external parameters are varied, these roots can collide and exchange places, giving rise to contour degenerations of the type we were able to exclude for affine-linear boundaries. Viewed globally, such collisions can (for instance) signal that two hypersurfaces have become tangent to one another, or that one of the hypersurfaces has become singular. Both of these phenomena already occur for the bubble integral.

The main difficulty is therefore no longer just identifying singular configurations in each integration-variable plane, but determining whether these configurations are globally compatible. For instance, verifying that a singular configuration in the $x_1$ plane that holds for fixed values of $x_2$ is compatible with one in the $x_2$ plane that holds for fixed values of $x_1$ requires tracking how each configuration moves as the other variable is varied. Moreover, compatible configurations need not be isolated points. Nevertheless, in Appendix~\ref{app:bubble} we illustrate how our strategy can be carried through for the massive bubble in the Baikov representation. Namely, we investigate the compatibility of singular configurations integration plane by integration plane, determine the contour deformations that are required to compute this integral's discontinuities, and show how sequences of discontinuities (together with the classification in Appendix~\ref{app:discontinuity_vs_branch_point_type}) constrain the symbol
even in the presence of algebraic roots. This suggests that our basic geometric strategy extends beyond hyperplane arrangements, although we leave the development of an algorithm that can be applied to more complicated Feynman integrals to future work.

\section{Conclusion}
\label{section:conclusion}

In this paper, we have explored how Landau analysis can be used to construct the symbol of twisted integrals that evaluate to multiple polylogarithms, either directly or order by order in an expansion around special values of the twist parameter. Our strategy involves classifying the singular configurations that give rise to logarithmic discontinuities, and computing each such discontinuity in the form of $2\pi i$ times an integral over a modified contour. Since these new integrals belong to the same class as the one we started with, this analysis can be applied recursively to construct the full symbol without ever evaluating the integral in terms of multiple polylogarithms.

We have worked this procedure out in full detail for twisted hyperplane integrals that involve simple poles. In this setting, the discontinuities that contribute at symbol level only come about via two elementary geometric moves---residue-type moves, which localize one integration variable, and twist-type moves, which introduce an additional power of $\epsilon$ without reducing the number of integrations. Twist factors that become independent of the remaining integration variables can likewise be treated using prefactor-type moves. Combining these operations recursively allows us to construct the weight-$(n{+}m)$ symbol at order $\epsilon^m$ by computing all sequences containing $n$ residue-type moves and $m$ twist- or prefactor-type moves. Notably, singularities that arise from pinch configurations need not be tracked, as they do not contribute at symbol level, and higher-multiplicity intersections can be accounted for by summing over the separate contributions from each participating twisted or singular hypersurface.

We have implemented this algorithm for twisted hyperplane integrals in an accompanying Mathematica notebook, which also includes a number of examples, several of which arise in the study of cosmological correlators. Although the code is intended as a proof of concept, the algorithm is already remarkably efficient---the examples it contains, including those highlighted in Section~\ref{section:algorithm}, run on a laptop in a fraction of the time required to integrate them into multiple polylogarithms (when the latter is feasible at all). We therefore expect our approach to be useful for exploring the symbols of cosmological correlators and related integrals at higher loop orders, and at higher orders in the twist expansion. 

Two limitations of our analysis deserve comment. First, the algorithm we presented targets the maximal-weight part of $\mathcal{I}^{(m)}(\mathbf{t})$, and branches of the recursion that do not allow all integrations to be evaluated in terms of residue contours are discarded, along with any lower-weight information they carry. Second, we have not sought to reconstruct contributions proportional to higher powers of $2\pi i$ that go beyond the symbol. Neither of these restrictions is intrinsic to our geometric approach, and it would be interesting to extend the analysis in Section~\ref{section:1dexample} to capture both classes of contributions. More generally, one could work out an analogous set of moves for generic values of the twist parameter, and try to reconstruct, from the same type of discontinuity information, the generalized hypergeometric functions that these integrals evaluate to.

A natural next step is to apply the same geometric strategy to compute the symbol of flat-space Feynman integrals from the Baikov representation. As discussed in Section~\ref{sec:beyond_hyperplanes}, this presents significant challenges, which we have begun to explore in Appendix~\ref{app:bubble}. Extending the full analysis to more complicated Feynman integrals will require a broader classification of singular configurations and explicit prescriptions for their difference contours. It will also require developing a more refined understanding of how sequences of discontinuities can be uplifted to symbol terms once algebraic branch points are present.

We expect that the geometric approach pursued here will be useful even where a complete algorithm is out of reach. Determining which sequences of discontinuities can arise in an integral yields constraints that can be leveraged in bootstrap approaches~\cite{Hannesdottir:2024hke}, even when the relative coefficients of different discontinuities cannot be computed directly. This offers an intermediate goal for integral families whose geometry is too complicated for the full construction to be carried out. More broadly, one of the core lessons of the hyperplane case---that the analytic structure of these integrals is encoded in a small number of elementary geometric moves---may well extend to more complicated classes of integrals. To what extent this remains true is, in our view, one of the central questions raised by this work.

\section*{Acknowledgements}

\noindent AJM is supported by the Royal Society grant URF{$\backslash$}R1{$\backslash$}221233, and additionally acknowledges support from the European Research Council (ERC) under the European Union’s Horizon Europe research and innovation program grant agreement 101163627 (ERC Starting Grant “AmpBoot”). AP is supported by the European Union (ERC, UNIVERSE PLUS, 101118787). LR is supported by the Royal Society via a Newton International Fellowship.

\appendix

\section{Branch Points Versus Discontinuities}
\label{app:discontinuity_vs_branch_point_type}

To construct the symbol of an integral from its discontinuities, one must discern which of these discontinuities produces a new factor of $2 \pi i$. However, the fact that a discontinuity generates $2 \pi i$ does not imply that the corresponding branch point is logarithmic (as discussed in Section~\ref{sec:beyond_hyperplanes}). In this appendix, we clarify our language around discontinuities versus branch points, and illustrate how longer sequences of discontinuities can be used to discern what type of branch point arises at a singular kinematic locus.

Following the main text, we refer to a discontinuity that is proportional to a single factor of $2\pi i$ as a logarithmic discontinuity. Branch points, by contrast, we classify according to how the monodromy operator $\mathcal{M}_\lambda$ acts under repeated application~\cite{pham2011singularities}. Writing 
\begin{equation}
\text{Disc}_\lambda = \mathcal{M}_\lambda - 1    
\end{equation} 
as in~\eqref{eq:disc_def}, iterated discontinuities can be computed by applying the operator
\begin{equation}
    \text{Disc}^c_\lambda = (\mathcal{M}_\lambda - 1)^c \, .
\end{equation}
Below, we highlight how two classes of branch points can be distinguished by the polynomial identity that $\mathcal{M}_\lambda$ satisfies. We make no attempt at a complete classification.

\vspace{.2cm}
\noindent {\bf Square-Root Branch Points}
\vspace{.1cm}

\noindent We say that $\lambda = 0$ is a square-root branch point of $I$ if $\mathcal{M}_\lambda$ acts nontrivially but with period two, namely
\begin{equation}
\mathcal{M}_{\lambda} I \neq I, \qquad \left( \mathcal{M}_{\lambda} \right)^2 I = I .
\label{eq:sqrt_type}
\end{equation}
This means that
\begin{align*}
(\mathcal{M}_\lambda - 1)^2 I & = (\mathcal{M}_\lambda^2 - 2\mathcal{M}_\lambda + 1) I \\
&= -2(\mathcal{M}_\lambda - 1) I\, ,
\end{align*}
and hence
\begin{equation}
\text{Disc}^c_{\lambda}(I) = (-2)^{c-1}\, \text{Disc}_{\lambda}(I) .
\label{eq:sqrt_iterated}
\end{equation}
A square-root branch point is therefore not characterized by its first discontinuity, but by the fact that repeated discontinuities return the same value (up to an integer multiple). We emphasize that this classification concerns only the action of $\mathcal{M}_\lambda$, and says nothing about whether or not $\text{Disc}_\lambda(I)$ is a logarithmic discontinuity, in the sense of being proportional to $2 \pi i$. 

\vspace{.2cm}
\noindent {\bf Logarithmic Branch Points}
\vspace{.1cm}

\noindent We say that $\lambda = 0$ is a logarithmic branch point of $I$ if repeated application of $\mathcal{M}_\lambda$ brings down the same term,
\begin{equation}
\mathcal{M}_{\lambda} I = I + I', \qquad \left(\mathcal{M}_{\lambda} \right)^c I = I + c\, I',
\label{eq:log_type}
\end{equation}
which requires $\mathcal{M}_{\lambda} I' = I'$. In this case $(\mathcal{M}_\lambda - 1)^2 I = 0$, so the second and all higher discontinuities vanish,
\begin{equation}
\text{Disc}_{\lambda}^2(I) = (\mathcal{M}^2_{\lambda} - 2\mathcal{M}_{\lambda} + 1) I = 0 .
\label{eq:log_iterated}
\end{equation}
The prototype is $I = \log\lambda$, for which $I' = 2\pi i$. More generally, a locus at which the letter $\lambda$ occupies the first $k$ entries of the symbol produces $(\mathcal{M}_\lambda - 1)^k I \neq 0$ but $(\mathcal{M}_\lambda - 1)^{k+1} I = 0$, so that exactly $k$ discontinuities can be computed before the sequence terminates. We include all such cases, in which $\mathcal{M}_\lambda - 1$ is nilpotent, under the heading of logarithmic branch points. These are precisely the branch points that appear as symbol letters.

The twist factors of Section~\ref{sec:hyperplane_integrals} furnish a compact example. A factor $\lambda^\epsilon$ satisfies
\begin{equation}
\left( \mathcal{M}_\lambda \right)^c \lambda^\epsilon = e^{2\pi i \epsilon c}\, \lambda^\epsilon = \sum_{m=0}^\infty \frac{\epsilon^m}{m!} \left( \log\lambda + 2\pi i\, c \right)^m ,
\label{eq:twist_monodromy}
\end{equation}
so that the coefficient of $\epsilon^m$ has a logarithmic branch point at $\lambda=0$, around which exactly $m$ discontinuities can be computed. Note that this statement only holds order by order in $\epsilon$; the resummed monodromy $e^{2\pi i\epsilon}$ is neither period two nor unipotent, and as such falls into neither of the above classes for generic $\epsilon$.

\section{The Bubble Feynman Integral}
\label{app:bubble}

In this appendix, we apply the type of plane-by-plane analysis discussed in Section~\ref{sec:hyperplane_integrals} to the massive bubble integral in the Baikov representation. Our aim is to identify the hypersurfaces in the space of external parameters that have nontrivial monodromy, and to determine whether each such hypersurface constitutes a logarithmic or square-root branch point. As anticipated in Section~\ref{sec:beyond_hyperplanes}, several of these singular hypersurfaces give rise to discontinuities that are linear in $2\pi i$ despite corresponding to square-root branch points. Nevertheless, the nature of the underlying branch point can still be determined by studying longer sequences of discontinuities.

Up to a numerical prefactor, the bubble integral in dimensional regularization can be written in Baikov representation as
\begin{equation}
I_{\rm bub} = (-s)^{1-\frac{D}{2}} \int_{\mathcal{B}\geq 0} \frac{dz_1}{(z_1 + m_1^2)}\frac{dz_2}{(z_2 + m_2^2)}\, \mathcal{B}^{\frac{D-3}{2}} \, ,
\label{eq:baikov_bubble}
\end{equation}
where the Baikov polynomial is
\begin{equation}
\mathcal{B}(z_1,z_2;s) := -s^2 - (z_1-z_2)^2 - 2s(z_1+z_2) \, .
\label{eq:baikov_poly}
\end{equation}
In the language of Section~\ref{sec:landau}, the propagators $(z_i + m_i^2)^{-1}$ define singular hypersurfaces, while $\mathcal{B} = 0$ is a twisted hypersurface that also serves as the integration boundary. For fixed $z_2$, the integration region $\mathcal{B} \geq 0$ in the $z_1$ plane corresponds to the interval between the two roots
\begin{equation}
z_1^\pm = -\left(\sqrt{s} \pm \sqrt{-z_2}\right)^2 \, ,
\label{eq:z1_roots}
\end{equation}
and similarly for $z_2$ at fixed values of $z_1$. Note that these roots of $\mathcal{B}$ are twisted endpoints in the sense of Section~\ref{section:1dexample}. Throughout, we take $D$ to be an integer, so that every branch point we encounter falls into one of the two classes of Appendix~\ref{app:discontinuity_vs_branch_point_type}.

As explained in Section~\ref{sec:hyperplane_integrals}, the integral $I_{\rm bub}$ can only become singular if a nontrivial singular configuration of hypersurfaces is realized in every integration-variable plane simultaneously. To test whether a candidate configuration meets this requirement, we fix $z_2$ at the location of a singular configuration and examine the $z_1$ plane, then similarly fix $z_1$ in a singular configuration and examine the $z_2$ plane. Only if both planes simultaneously exhibit a nontrivial intersection can the integral develop a singularity; we then say that the two configurations are compatible.

There are three ways in which the twisted and singular hypersurfaces can intersect. The Baikov polynomial can become degenerate on its own, it can intersect one of the propagators, or it can intersect both. We now study each of these situations in turn.

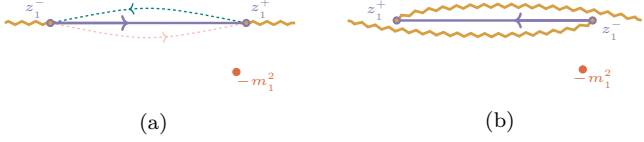
\begin{figure}
    \centering
    \subfloat[\label{subfig:bubble_s=0_a}]{%
        \begin{minipage}[c]{0.47\columnwidth}
            \centering
            \resizebox{\linewidth}{!}{%
                \input{tikz_figs/bubble/s=0_a}%
            }
        \end{minipage}%
    }
    \hfill
    \subfloat[\label{subfig:bubble_s=0_b}]{%
        \begin{minipage}[c]{0.47\columnwidth}
            \centering
            \resizebox{\linewidth}{!}{%
                \input{tikz_figs/bubble/s=0_b}%
            }
        \end{minipage}%
    }
    \caption{Tracking the contour in the complex plane of $z_1$ as $s$ is continued counterclockwise about zero. (a) The set up before continuation with dashed paths showing the trajectory of the endpoints as $s$ is continued. (b) After $s$ completes a full circuit about the origin, the roots have traded place and the contour is moved onto the adjacent Riemann sheet.}
    \label{fig:bubble_s=0}
\end{figure}

\vspace{.2cm}
\noindent {\bf No Propagators On Shell}
\vspace{.1cm}

\noindent Since $\mathcal{B}$ is quadratic in ${z_1 - z_2}$ but only linear in ${z_1 + z_2}$, the curve $\mathcal{B} = 0$ is a parabola with axis along the line ${z_1 = z_2}$. At ${s = 0}$ it degenerates to the double line ${\mathcal{B} = -(z_1 - z_2)^2}$. In the $z_1$ plane, this manifests as the roots~\eqref{eq:z1_roots} coinciding at $z_1 = z_2$ for every value of $z_2$. Continuing $s$ counterclockwise around zero causes the two roots to exchange places, so that the contour reverses orientation and moves onto an adjacent Riemann sheet, as seen in Figure~\ref{fig:bubble_s=0}. The same thing happens in the $z_2$ plane, since $\mathcal{B}$ is symmetric under $z_1 \leftrightarrow z_2$. This means that the two configurations are compatible not just at an isolated point, but along the entire line $z_1 = z_2$.

Under this analytic continuation around $s=0$, the integral picks up a factor of $(-1)e^{2\pi i (D-3)/2}$, where the overall sign comes from the reversed orientation, and the phase from the change of sheet.\footnote{More specifically, the factor of $-1$ is the one-dimensional case of the general rule $(-1)^n$ for $n$ independent transverse pinches. This rule concerns only the orientation of the contour and does not include the accompanying phase.} This can also be derived via a change of variables similar to the one we used to analyze pairs of twisted branch points in Section~\ref{section:1dexample}. At fixed $z_2$, we can write $\mathcal{B} = (z_1^- - z_1)(z_1 - z_1^+)$, so that both endpoints of the $z_1$ integration are twisted with exponent $(D-3)/2$. Changing variables to $z_1 = z_1^+ + (z_1^- - z_1^+)\,t$, the $z_1$ integral in~\eqref{eq:baikov_bubble} becomes
\begin{equation}
\begin{gathered}
\left(z_1^- - z_1^+\right)^{D-2}
\int_0^1 dt\,
\frac{\left[t(1-t)\right]^{\frac{D-3}{2}}}
{z_1(t)+m_1^2} \, ,
\\[3pt]
z_1^- - z_1^+
=
4\sqrt{s}\sqrt{-z_2} \, .
\end{gathered}
\label{eq:s_zero_cov}
\end{equation}
Continuing $s$ once around zero sends $\sqrt{s} \to -\sqrt{s}$, under which the prefactor picks up $e^{i\pi(D-2)} = (-1)\,e^{2\pi i(D-3)/2}$, while the remaining integral is invariant (as the exchange $z_1^\pm \to z_1^\mp$ is compensated by $t \to 1-t$). However, the prefactor $(-s)^{1-D/2}$ in~\eqref{eq:baikov_bubble} simultaneously picks up a factor of $e^{2\pi i(1-D/2)}$, which cancels the phase generated by the integral. The monodromy of $I_{\rm bub}$ about $s = 0$ is therefore trivial in all dimensions (for the physical integration contour), so we have $\text{Disc}_s(I_{\rm bub}) = 0$.

\vspace{.2cm}
\noindent {\bf One Propagator On Shell}
\vspace{.1cm}

\noindent We next consider the situation in which $\mathcal{B} = 0$ meets the first propagator hypersurface $z_1 = -m_1^2$ but not the second. For generic parameters, the line $z_1 = -m_1^2$ crosses the parabola $\mathcal{B} = 0$ at two points, $z_2 = -(\sqrt{s} \pm m_1)^2$, so in the $z_1$ plane the pole sits on an endpoint for these two values of $z_2$. However, neither crossing is by itself singular---a nontrivial intersection must also occur in the $z_2$ plane. Having stipulated that the pole at $z_2 = -m_2^2$ is not involved, the only possibility is that the endpoints $z_2^\pm = -(\sqrt{s} \pm \sqrt{-z_1})^2$ collide. This happens only if $s = 0$ (the situation already treated above), or if $z_1 = 0$. In the latter case, the line $z_1 = 0$ is tangent to the parabola at $(z_1, z_2) = (0,-s)$; the relevant singularity therefore occurs at $m_1^2 = 0$, where the propagator hyperplane becomes this tangent line and its two crossings with $\mathcal{B} = 0$ merge into one.

Fixing $z_2 = -s$ and continuing $m_1^2$ around the origin, the pole at $z_1 = -m_1^2$ encircles the endpoint $z_1^- = 0$ and picks up a residue contour, while in the $z_2$ plane the endpoints $z_2^\pm = -(\sqrt{s} \pm m_1)^2$ exchange places, reversing the orientation of the contour and moving it onto an adjacent Riemann sheet. Applying this analytic continuation $c$ times, we find
\begin{equation}
\begin{aligned}
\mathcal{M}^c_{m_1^2}\, I_{\rm bub}
&=
I_{\rm bub}
+
2\pi i
\left(
\sum_{j=1}^{c} e^{j\pi i D}
\right)
(-s)^{1-\frac{D}{2}}
\\
&\ \ \times
\int_{\mathcal{B}\geq 0}
\frac{dz_2}{z_2+m_2^2}
\left[
\mathcal{B}(-m_1^2,z_2;s)
\right]^{\frac{D-3}{2}} \, ,
\end{aligned}
\label{eq:M_m1_iterated}
\end{equation}
where the phase $e^{i\pi D}$ is the product of the orientation reversal $(-1)$ and the sheet change $e^{i\pi(D-3)}$. In even dimensions, this phase evaluates to 1 and $m_1^2 = 0$ corresponds to a logarithmic branch point, while in odd dimensions this singular hypersurface is a square-root branch point. This is consistent with our expectations~\cite{Hannesdottir:2021kpd}.

\begin{figure} 
    \centering
    \subfloat[\label{subfig:bubble_T_a}]{%
        \begin{minipage}[c]{0.47\columnwidth}
            \centering
            \resizebox{\linewidth}{!}{%
                \input{tikz_figs/bubble/T_a}%
            }
        \end{minipage}%
    }
    \hfill
    \subfloat[\label{subfig:bubble_T_b}]{%
        \begin{minipage}[c]{0.47\columnwidth}
            \centering
            \resizebox{\linewidth}{!}{%
                \input{tikz_figs/bubble/T_b}%
            }
        \end{minipage}%
    }
    \caption{Contour deformation in the complex plane of $z_1$ with $z_2$ fixed. (a) The initial arrangement with the path of the relevant endpoint shown by the dashed line. (b) The deformed contour after $s$ is continued in a complete circuit around the threshold.}
    \label{fig:bubble_T}
\end{figure}
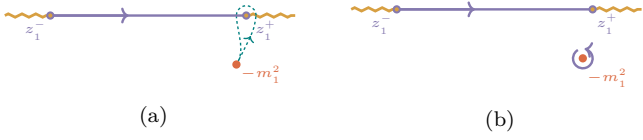

\vspace{.2cm}
\noindent {\bf Both Propagators On Shell}
\vspace{.1cm}

\noindent Finally, we consider the situation in which $\mathcal{B} = 0$ meets both propagator hypersurfaces at once. This occurs when $\mathcal{B}$ vanishes at the point $(z_1, z_2) = (-m_1^2, -m_2^2)$. Recalling the notation $r_\pm=m_1\pm m_2$ from Section~\ref{sec:beyond_hyperplanes}, we have
\begin{equation}
\mathcal{B}(-m_1^2,-m_2^2;s) = \big(s-r_+^2\big)\big(r_-^2 - s\big) \, ,
\label{eq:B_at_poles}
\end{equation}
so that this happens at the threshold $s = r_+^2$ and at the pseudo-threshold $s = r_-^2$. However, we still must check whether the singular configurations in the two planes are compatible at each of these locations.

We check the threshold first. Fixing $z_2 = -m_2^2$, the roots in the $z_1$ plane become $z_1^\pm = -(\sqrt{s} \pm m_2)^2$, and the upper endpoint $z_1^-$ coincides with the pole at ${z_1 = -m_1^2}$ precisely when $s = r_+^2$. Continuing $s$ around this threshold value, the pole encircles this endpoint and picks up a residue contour, as shown in Figure~\ref{fig:bubble_T}. We see that this is a residue-type move. If we instead fix $z_1 = -m_1^2$, the same thing happens in the $z_2$ plane, with the pole at $z_2 = -m_2^2$ encircling the upper endpoint $z_2^-$. Since both poles encircle the upper endpoint in their respective planes, the two residue contours lie on the same Riemann sheet, and combine into a difference contour that localizes the integral at $(z_1,z_2) = (-m_1^2,-m_2^2)$. The resulting discontinuity is
\begin{equation}
\begin{aligned}
\text{Disc}_{s-r_+^2}(I_{\rm bub})
&=
(2\pi i)^2\,(-s)^{1-\frac{D}{2}}
\\
&\qquad\times
\left[
\mathcal{B}(-m_1^2,-m_2^2;s)
\right]^{\frac{D-3}{2}} \, ,
\end{aligned}
\label{eq:disc_T}
\end{equation}
which is linear in $2\pi i$ (relative to the normalization seen in~\eqref{eq:massive_bubble_2d}).

At the pseudo-threshold, by contrast, the configurations in the $z_1$ and $z_2$ plane are not compatible. Taking $m_1 > m_2$, one finds that at $s = r_-^2$ the pole in the $z_1$ plane coincides with the lower endpoint $z_1^+$, while the pole in the $z_2$ plane coincides with the upper endpoint $z_2^-$. This means that the copies of the poles that deform the contour (if we track them as in Figure~\ref{fig:1D_branch_choice}) lie on adjacent Riemann sheets, allowing for the continued contour to be deformed back to the original contour. The monodromy of $I_{\rm bub}$ about $s = r_-^2$ is thus trivial, even though $\mathcal{B}$ meets both poles there.

Repeated analytic continuation around the threshold $s=r_+^2$ acts on~\eqref{eq:disc_T} through the factor $\smash{(s-r_+^2)^{(D-3)/2}}$, each loop contributing a phase $\smash{e^{i\pi(D-3)}}$. Applying this analytic continuation $c$ times, we find
\begin{equation}
\begin{aligned}
\mathcal{M}_{s-r_+^2}^c\, I_{\rm bub}
&=
I_{\rm bub}
+
(2\pi i)^2
\left(
\sum_{j=1}^{c} e^{(j-1) i\pi (D-3)}
\right)
\\
&\ \ \ \times (-s)^{1-\frac{D}{2}} 
\left[\mathcal{B}(-m_1^2,-m_2^2;s)\right]^{\frac{D-3}{2}} .
\end{aligned}
\label{eq:MT_iterated}
\end{equation}
In odd dimensions the phase $e^{i\pi(D-3)}$ is unity, so the sum equals $c$ and~\eqref{eq:MT_iterated} takes the form $\smash{\mathcal{M}_{s-r_+^2}^c\, I_{\rm bub} = I_{\rm bub} + c\, I'}$ with $\smash{I' = \text{Disc}_{s-r_+^2}(I_{\rm bub})}$; this makes it a logarithmic branch point. In even dimensions the phase is $-1$, so the sum is $1$ for odd $c$ and $0$ for even $c$. In particular, $\smash{\mathcal{M}_{s-r_+^2}\, I_{\rm bub} = I_{\rm bub} + \text{Disc}_{s-r_+^2}(I_{\rm bub})}$ but $\smash{\mathcal{M}_{s-r_+^2}^2\, I_{\rm bub} = I_{\rm bub}}$---the monodromy acts nontrivially but with period two, making this a square-root branch point. Note that the parity dependence is opposite to that of~\eqref{eq:M_m1_iterated}, since the residue-type moves that localize the integral here involve no reversal of orientation. In particular, in $D = 2$ the threshold produces a discontinuity linear in $2\pi i$, even though the branch point exhibits square-root behavior, as described in Section~\ref{sec:beyond_hyperplanes}.

We can also ask what further branch points the threshold discontinuity~\eqref{eq:disc_T} possesses. Since all integrals have been evaluated, this discontinuity is an algebraic function of $s$ whose only branch points are at $s = 0$, $r_+^2$, and $r_-^2$. We have already treated the threshold case, but for the remaining two we find that
\begin{equation}
\begin{aligned}
\mathcal{M}^c_s\, \text{Disc}_{s-r_+^2}(I_{\rm bub}) &= e^{-c\, i\pi D}\, \text{Disc}_{s-r_+^2}(I_{\rm bub}) \, , \\
\mathcal{M}^c_{s-r_-^2}\, \text{Disc}_{s-r_+^2}(I_{\rm bub}) &= e^{c\, i\pi (D-3)}\, \text{Disc}_{s-r_+^2}(I_{\rm bub}) \, .
\end{aligned}
\label{eq:disc_T_monodromies}
\end{equation}
In every integer dimension these phases are $\pm 1$, so all subsequent branch points of the threshold discontinuity are trivial or of square-root type, and no further logarithmic discontinuities can be computed.

\vspace{.2cm}
\noindent {\bf Summary}
\vspace{.1cm}

\noindent On its principal sheet, the bubble integral has nontrivial monodromies about the singular hypersurfaces $s = r_+^2$, $m_1^2 = 0$, and $m_2^2 = 0$, while the monodromies about $s = 0$ and $s = r_-^2$ only become nontrivial after a discontinuity has been computed. All three nontrivial monodromies produce discontinuities linear in $2\pi i$, so tracking factors of $2\pi i$ alone might suggest that all three singular hypersurfaces give rise to symbol letters. However, the branch-point classification says otherwise. In even dimensions only $m_1^2 = 0$ and $m_2^2 = 0$ correspond to logarithmic branch points, while in odd dimensions only the threshold is a logarithmic branch point. This is consistent with what is found in both $D=2$ and $D=3$~\cite{Hannesdottir:2024hke}. Thus, when algebraic branch points are present, the factors of $2\pi i$ generated by a discontinuity do not by themselves determine whether it corresponds to a symbol letter; longer sequences of discontinuities must be computed to discern what type of branch point a given singular hypersurface constitutes.

\bibliographystyle{apsrev4-2}
\bibliography{refs}

\end{document}

%% file: tikz_figs/1dim/2a.tex
\begin{tikzpicture}

    \definecolor{customred}{RGB}{219,107,67}
    \definecolor{custompurple}{RGB}{133,121,175}
    \definecolor{customyellow}{RGB}{215,159,65}


        \draw[
        teal,
        line width=0.4pt,
        dash pattern=on 0.8pt off 0.8pt
    ] (2,0.1) arc[
        start angle=90,
        end angle=0,
        radius=0.1
    ];
        \draw[
        teal,
        line width=0.4pt,
        dash pattern=on 0.8pt off 0.8pt
    ] (2,0.1) arc[
        start angle=90,
        end angle=180,
        radius=0.1
    ];
    
    \draw[
        teal,
        line width=0.4pt,
        dash pattern=on 0.8pt off 0.8pt,
        decoration={
            markings,
            mark=at position 0.8 with {
                \arrow{>[scale=0.65]}
            }
        },
        postaction={decorate}
    ] (1.4,-0.5) .. controls (2.0,-0.2) .. (2.1,0);

    \draw[
        teal,
        line width=0.4pt,
        dash pattern=on 0.8pt off 0.8pt,
    ] (1.4,-0.5) .. controls (1.8,-0.2) .. (1.9,0);

    \draw [line width=0.7pt,color=custompurple,->,-{To[scale=0.7]}] (0,0) -- (0.8,0);
    \draw [line width=0.7pt,color=custompurple] (0,0) -- (2,0);

    \node[font=\fontsize{4}{12},custompurple] at (-0.1,0.2) {$b_1$};
    \node[font=\fontsize{4}{12},custompurple] at (2.1,0.2) {$b_2$};

    \fill[custompurple] (0,0)   circle (1.2pt);
    \fill[custompurple] (2,0)   circle (1.2pt);

    \fill[customred] (1.4,-0.5) circle (1.2pt);
    \node [font=\fontsize{4}{12},customred] at (1.2,-0.6) {$a_k$};
\end{tikzpicture}

%% file: tikz_figs/1dim/2b.tex
\begin{tikzpicture}

    \definecolor{customred}{RGB}{219,107,67}
    \definecolor{custompurple}{RGB}{133,121,175}
    \definecolor{customyellow}{RGB}{215,159,65}


\draw[
    line width=0.7pt,
    color=custompurple,
    line cap=round,
    line join=round
]
    (0,0) -- (1.7,0) .. controls (1.8,0) .. (1.7,-0.1) -- (1.35,-0.43) .. controls (1.25,-0.53) and (1.36,-0.65) .. (1.46,-0.55) -- (2,0);

    \draw[
        line width=0.7pt,
        color=custompurple,
        -{To[scale=0.7]}
    ] (0,0) -- (0.8,0);

    \fill[custompurple] (0,0) circle (1.2pt);
    \fill[custompurple] (2,0) circle (1.2pt);
    
    \node[font=\fontsize{4}{12},custompurple] at (-0.1,0.2) {$b_1$};
    \node[font=\fontsize{4}{12},custompurple] at (2.1,0.2) {$b_2$};

    \fill[customred] (1.4,-0.5) circle (1.2pt);
    \node [font=\fontsize{4}{12},customred] at (1.2,-0.6) {$a_k$};

\end{tikzpicture}

%% file: tikz_figs/1dim/1a.tex
\begin{tikzpicture}

    \definecolor{customred}{RGB}{219,107,67}
    \definecolor{custompurple}{RGB}{133,121,175}
    \definecolor{customyellow}{RGB}{215,159,65}


    \draw[
        teal,
        line width=0.4pt,
        dash pattern=on 0.8pt off 0.8pt
    ] (2,0.1) arc[
        start angle=90,
        end angle=0,
        radius=0.1
    ];
        \draw[
        teal,
        line width=0.4pt,
        dash pattern=on 0.8pt off 0.8pt
    ] (2,0.1) arc[
        start angle=90,
        end angle=180,
        radius=0.1
    ];
    
    \draw[
        teal,
        line width=0.4pt,
        dash pattern=on 0.8pt off 0.8pt,
        decoration={
            markings,
            mark=at position 0.8 with {
                \arrow{>[scale=0.65]}
            }
        },
        postaction={decorate}
    ] (1.4,-0.5) .. controls (2.0,-0.2) .. (2.1,0);

    \draw[
        teal,
        line width=0.4pt,
        dash pattern=on 0.8pt off 0.8pt,
    ] (1.4,-0.5) .. controls (1.8,-0.2) .. (1.9,0);

    \draw[
        line width=0.7pt,
        color=custompurple,
        -{To[scale=0.7]}
    ] (0,0) -- (0.8,0);

    \draw[
        line width=0.7pt,
        color=custompurple
    ] (0,0) -- (2,0);

    \node[font=\fontsize{4}{12},custompurple] at (-0.1,0.2) {$b_1$};
    \node[font=\fontsize{4}{12},custompurple] at (2.1,0.2) {$b_2$};

    \fill[custompurple] (0,0) circle (1.2pt);
    \fill[custompurple] (2,0) circle (1.2pt);

    \fill[customyellow] (1.4,-0.5) circle (1.2pt);
    \node[
        font=\fontsize{4}{12},
        customyellow
    ] at (1.2,-0.6) {$s_k$};

    \draw[
        thick,
        customyellow,
        decoration={
            zigzag,
            amplitude=0.5pt,
            segment length=2pt,
            pre=moveto,
            pre length=1pt
        },
        decorate
    ] (1.4,-0.5) -- (1.20912,-1);

\end{tikzpicture}

%% file: tikz_figs/1dim/1b.tex
\begin{tikzpicture}

    \definecolor{customred}{RGB}{219,107,67}
    \definecolor{custompurple}{RGB}{133,121,175}
    \definecolor{customyellow}{RGB}{215,159,65}


\draw[
    line width=0.7pt,
    color=custompurple,
    line cap=round,
    line join=round
]
    (0,0) -- (1.7,0) .. controls (1.8,0) .. (1.7,-0.1) -- (1.35,-0.43) .. controls (1.25,-0.53) and (1.36,-0.65) .. (1.46,-0.55) -- (2,0);

    \draw[
        line width=0.7pt,
        color=custompurple,
        -{To[scale=0.7]}
    ] (0,0) -- (0.8,0);

    \fill[custompurple] (0,0) circle (1.2pt);
    \fill[custompurple] (2,0) circle (1.2pt);
    
    \node[font=\fontsize{4}{12},custompurple] at (-0.1,0.2) {$b_1$};
    \node[font=\fontsize{4}{12},custompurple] at (2.1,0.2) {$b_2$};

    \fill[customyellow] (1.4,-0.5) circle (1.2pt);
    \node [font=\fontsize{4}{12},customyellow] at (1.2,-0.6) {$s_k$};
    \draw[
        line width=0.5pt,
        customyellow,
        decoration={
            zigzag,
            amplitude=0.5pt,
            segment length=2pt,
            pre=moveto,
            pre length=1pt
        },
        decorate
    ]
        (1.4,-0.5) -- (1.92667, 0.0679931) .. controls (1.99466,0.141321) and (2.14132, 0.00533429) .. (2.07333,-0.0679931) -- (1.20912,-1);

\end{tikzpicture}

%% file: tikz_figs/1dim/twisted_enpoint_riemann_sheets_x1.5.tex
\tdplotsetmaincoords{68}{118}

\begin{tikzpicture}[
    /pgf/fpu/install only={reciprocal},
    tdplot_main_coords,
    line cap=round,
    line join=round
]


\definecolor{customred}{RGB}{219,107,67}
\definecolor{custompurple}{RGB}{133,121,175}
\definecolor{customyellow}{RGB}{215,159,65}
\definecolor{sheetcolor}{RGB}{252,252,245}
\definecolor{sheetedge}{RGB}{105,101,122}

\def\rmin{0.32}
\def\rmax{3.35}

\def\pitch{1.35}
\def\dtheta{6}

\def\rpath{2.35}

\def\thetastart{70}
\def\thetaend{435}
\def\thetaextra{-295}

\def\thetamin{-360}
\def\thetamax{660}

\def\thetacut{18}

\def\rcontourend{3}
\def\thetacontourend{30}

\def\thetaarrowmid{100}
\def\arrowhalfwidth{12}

%

\foreach \t in {-360,-354,...,654} {

    \pgfmathsetmacro{\tp}{\t+\dtheta}

    \path[
        fill=sheetcolor,
        draw=none,
        opacity=.96
    ]
        ({\rmin*cos(\t)},
         {\rmin*sin(\t)},
         {\pitch*\t/360})
    --
        ({\rmax*cos(\t)},
         {\rmax*sin(\t)},
         {\pitch*\t/360})
    --
        ({\rmax*cos(\tp)},
         {\rmax*sin(\tp)},
         {\pitch*\tp/360})
    --
        ({\rmin*cos(\tp)},
         {\rmin*sin(\tp)},
         {\pitch*\tp/360})
    -- cycle;
}


\foreach \t in {-360,-336,...,648} {
    \draw[
        white,
        line width=0.18pt,
        opacity=.18
    ]
        ({\rmin*cos(\t)},
         {\rmin*sin(\t)},
         {\pitch*\t/360})
    --
        ({\rmax*cos(\t)},
         {\rmax*sin(\t)},
         {\pitch*\t/360});
}


\draw[
    sheetedge!60,
    line width=0.25pt,
    samples=240,
    smooth,
    variable=\t,
    domain=\thetamin:\thetamax
]
plot
(
    {\rmin*cos(\t)},
    {\rmin*sin(\t)},
    {\pitch*\t/360}
);


\draw[
    sheetedge!35,
    line width=0.25pt,
    samples=240,
    smooth,
    variable=\t,
    domain=\thetamin:\thetamax
]
plot
(
    {\rmax*cos(\t)},
    {\rmax*sin(\t)},
    {\pitch*\t/360}
);


\newcommand{\visiblepath}[2]{%
    \draw[
        teal,
        line width=1.1pt,
        dash pattern=on 2pt off 3.17pt,
        samples=140,
        smooth,
        variable=\t,
        domain=#1:#2
    ]
    plot
    (
        {\rpath*cos(\t)},
        {\rpath*sin(\t)},
        {\pitch*\t/360}
    );
}

\newcommand{\visiblepatharrow}[2]{%
    \draw[
        teal,
        line width=1.1pt,
        dash pattern=on 2pt off 3.17pt,
        samples=140,
        smooth,
        variable=\t,
        domain=#1:#2,
        decoration={
            markings,
            mark=at position 0.5 with {
                \arrow{>[scale=0.975]}
            }
        },
        postaction={decorate}
    ]
    plot
    (
        {\rpath*cos(\t)},
        {\rpath*sin(\t)},
        {\pitch*\t/360}
    );
}

%
%

\visiblepath{-360}{-272}
\visiblepatharrow{-272}{-248}
\visiblepath{-248}{-233}
\visiblepath{-99}{0}

\visiblepath{0}{88}
\visiblepatharrow{88}{112}
\visiblepath{112}{127}
\visiblepath{201}{360}

\visiblepath{360}{448}
\visiblepatharrow{448}{472}
\visiblepath{472}{660}


\coordinate (akmiddle) at
(
    {\rpath*cos(\thetastart)},
    {\rpath*sin(\thetastart)},
    {\pitch*\thetastart/360}
);

\coordinate (aktop) at
(
    {\rpath*cos(\thetaend)},
    {\rpath*sin(\thetaend)},
    {\pitch*\thetaend/360}
);

\coordinate (akbottom) at
(
    {\rpath*cos(\thetaextra)},
    {\rpath*sin(\thetaextra)},
    {\pitch*\thetaextra/360}
);

\fill[customred] (akmiddle) circle (1.8pt);
\fill[customred] (aktop)    circle (1.8pt);
\fill[customred] (akbottom) circle (1.8pt);

\node[
    customred,
    font=\fontsize{10}{12}\selectfont,
    scale=0.8,
    transform shape,
    anchor=west,
    xshift=-16pt,
    yshift=5pt
]
at (akmiddle) {$a_k$};

\node[
    customred,
    font=\fontsize{10}{12}\selectfont,
    scale=0.8,
    transform shape,
    anchor=west,
    xshift=-16pt,
    yshift=5pt
]
at (aktop) {$a_k$};

\node[
    customred,
    font=\fontsize{10}{12}\selectfont,
    scale=0.8,
    transform shape,
    anchor=east,
    xshift=1pt,
    yshift=5pt
]
at (akbottom) {$a_k$};

%

\draw[
    line width=1.05pt,
    color=custompurple,
    decoration={
        markings,
        mark=at position 0.5 with {
            \arrow{>[scale=1.05]}
        }
    },
    postaction={decorate}
]
    ({\rcontourend*cos(\thetacontourend)},
     {\rcontourend*sin(\thetacontourend)},
     {\pitch*\thetacontourend/360})
--
        ({0.87*cos(\thetacut)},
     {0.24*sin(\thetacut)},
     {\pitch*\thetacut/360});

\coordinate (branchpoint) at
    ({0.87*cos(\thetacut)},
     {0.24*sin(\thetacut)},
     {\pitch*\thetacut/360});

\coordinate (contourendpoint) at
    ({\rcontourend*cos(\thetacontourend)},
     {\rcontourend*sin(\thetacontourend)},
     {\pitch*\thetacontourend/360});
     
\fill[custompurple] (branchpoint) circle (1.8pt);
\fill[custompurple] (contourendpoint) circle (1.8pt);

%
\node[
    custompurple,
    font=\fontsize{10}{12}\selectfont,
    scale=0.8,
    transform shape,
    anchor=south east,
    xshift=1pt,
    yshift=-1pt
]
    at (branchpoint) {$b_2$};

\node[
    custompurple,
    font=\fontsize{10}{12}\selectfont,
    scale=0.8,
    transform shape,
    anchor=south west,
    xshift=-16pt,
    yshift=-7pt
]
    at (contourendpoint) {$b_1$};


\newcommand{\branchcut}[1]{%
    \draw[
        color=customyellow,
        line width=1.05pt,
        decorate,
        decoration={
            zigzag,
            amplitude=1.2pt,
            segment length=4pt
        }
    ]
        ({0.87*cos(#1)},
         {0.24*sin(#1)},
         {\pitch*#1/360})
    --
        ({3.66*cos(#1)},
         {0.3*sin(#1)},
         {\pitch*#1/360});
}

\branchcut{-342}
\branchcut{18}
\branchcut{378}

\end{tikzpicture}

%% file: tikz_figs/1dim/3a.tex
\begin{tikzpicture}

    \definecolor{customred}{RGB}{219,107,67}
    \definecolor{custompurple}{RGB}{133,121,175}
    \definecolor{customyellow}{RGB}{215,159,65}


        \draw[
        teal,
        line width=0.4pt,
        dash pattern=on 0.8pt off 0.8pt
    ] (1.1,0.) arc[
        start angle=0,
        end angle=150,
        radius=0.1
    ];
    
    \draw[
        teal,
        line width=0.4pt,
        dash pattern=on 0.8pt off 0.8pt,
        decoration={
            markings,
            mark=at position 0.6 with {
                \arrow{>[scale=0.65]}
            }
        },
        postaction={decorate}
    ] (1.,-0.5) .. controls (1.1,-0.2) .. (1.1,0);

    \draw[
        teal,
        line width=0.4pt,
        dash pattern=on 0.8pt off 0.8pt,
    ] (1.,-0.5) .. controls (.9,-0.2) .. (.9,-.05);

    \draw[
        pink,
        line width=0.4pt,
        dash pattern=on 0.8pt off 0.8pt
    ] (0.9,0) arc[
        start angle=-180,
        end angle=-20,
        radius=0.1
    ];
    
    \draw[
        pink,
        line width=0.4pt,
        dash pattern=on 0.8pt off 0.8pt,
        decoration={
            markings,
            mark=at position 0.6 with {
                \arrow{>[scale=0.65]}
            }
        },
        postaction={decorate}
    ] (1.,0.5) .. controls (.9,0.2) .. (.9,0);

    \draw[
        pink,
        line width=0.4pt,
        dash pattern=on 0.8pt off 0.8pt,
    ] (1.,0.5) .. controls (1.1,0.2) .. (1.1,.05);

    \draw[
        line width=0.7pt,
        color=custompurple,
        -{To[scale=0.7]}
    ] (0,0) -- (0.8,0);

    \draw[
        line width=0.7pt,
        color=custompurple
    ] (0,0) -- (2,0);

    \node[font=\fontsize{4}{12},custompurple] at (-0.1,0.2) {$b_1$};
    \node[font=\fontsize{4}{12},custompurple] at (2.1,0.2) {$b_2$};

    \fill[custompurple] (0,0) circle (1.2pt);
    \fill[custompurple] (2,0) circle (1.2pt);

    \fill[customyellow] (1,-0.5) circle (1.2pt);
    \node[
        font=\fontsize{4}{12},
        customyellow
    ] at (0.7,-0.6) {$s_1$};

    \draw[
        thick,
        customyellow,
        decoration={
            zigzag,
            amplitude=0.5pt,
            segment length=2pt,
            pre=moveto,
            pre length=1pt
        },
        decorate
    ] (1.,-0.5) -- (1.,-1);

    \fill[customyellow] (1,0.5) circle (1.2pt);
    \node[
        font=\fontsize{4}{12},
        customyellow
    ] at (1.3,0.6) {$s_2$};

    \draw[
        thick,
        customyellow,
        decoration={
            zigzag,
            amplitude=0.5pt,
            segment length=2pt,
            pre=moveto,
            pre length=1pt
        },
        decorate
    ] (1.,0.5) -- (1.,1);

\end{tikzpicture}

%% file: tikz_figs/1dim/3b.tex
\begin{tikzpicture}[/pgf/fpu/install only={reciprocal}]

    \definecolor{customred}{RGB}{219,107,67}
    \definecolor{custompurple}{RGB}{133,121,175}
    \definecolor{customyellow}{RGB}{215,159,65}


    \draw[
        color=custompurple,
        line width=0.7pt,
    ] (1.1,0.5) arc[
        start angle=0,
        end angle=180,
        radius=0.1
    ];

        \draw[
        color=custompurple,
        line width=0.7pt,
    ] (1.2,0.5) arc[
        start angle=0,
        end angle=180,
        radius=0.2
    ];

        \draw[
        color=custompurple,
        line width=0.7pt,
    ] (1.1,-0.5) arc[
        start angle=0,
        end angle=-180,
        radius=0.1
    ];

        \draw[
        color=custompurple,
        line width=0.7pt,
    ] (1.2,-0.5) arc[
        start angle=0,
        end angle=-180,
        radius=0.2
    ];

    \draw[
        line width=0.7pt,
        color=custompurple,
        -{To[scale=0.7]}
    ] (0,0) -- (0.4,0);

    \draw[
        line width=0.7pt,
        color=custompurple
    ] (0,0) -- (.6,0);

    \draw[
        line width=0.7pt,
        color=custompurple,
        -{To[scale=0.7]}
    ] (1.4,0) -- (1.8,0);

    \draw[
        line width=0.7pt,
        color=custompurple
    ] (1.4,0) -- (2,0);

    \draw[
        line width=0.7pt,
        color=custompurple
    ] (0.6,0) .. controls (0.75,0) .. (.8,-0.5);

    \draw[
        line width=0.7pt,
        color=custompurple
    ] (1.2,0.5) .. controls (1.25,0) .. (1.4,0);

    \draw[
        line width=0.7pt,
        color=custompurple
        ] (1.2,-0.5) .. controls (1.2,-.25) .. (1.15,0.0) .. controls (1.1,0.25) .. (1.1,0.5);

    \draw[
        line width=0.7pt,
        color=custompurple
        ] (1.1,-0.5) .. controls (1.1,-.25) .. (1.,0.0) .. controls (0.9,0.25) .. (0.9,0.5);

    \draw[
        line width=0.7pt,
        color=custompurple
        ] (0.9,-0.5) .. controls (0.9,-.25) .. (.85,0.0) .. controls (0.8,0.25) .. (0.8,0.5);

    \node[font=\fontsize{4}{12},custompurple] at (-0.1,0.2) {$b_1$};
    \node[font=\fontsize{4}{12},custompurple] at (2.1,0.2) {$b_2$};

    \fill[custompurple] (0,0) circle (1.2pt);
    \fill[custompurple] (2,0) circle (1.2pt);

    \fill[customyellow] (1,-0.5) circle (1.2pt);
    \node[
        font=\fontsize{4}{12},
        customyellow
    ] at (0.7,-0.6) {$s_1$};

    \draw[
    thick,
    customyellow,
    decoration={
        zigzag,
        amplitude=0.5pt,
        segment length=2pt,
        pre=moveto,
        pre length=1pt
    },
    decorate
]
    (1,-0.5) -- (0.925,0) .. controls (0.85,0.25) .. (0.85,0.5)
    arc[
        start angle=180,
        end angle=0,
        radius=0.15
    ]
    .. controls (1.15,0.25) .. (1.2,0.0) .. controls (1.3,-0.5) .. (1,-1);

    \fill[customyellow] (1,0.5) circle (1.2pt);
    \node[
        font=\fontsize{4}{12},
        customyellow
    ] at (1.3,0.6) {$s_2$};

    \draw[
    thick,
    customyellow,
    decoration={
        zigzag,
        amplitude=0.5pt,
        segment length=2pt,
        pre=moveto,
        pre length=1pt
    },
    decorate
]
    (1,0.5) -- (1.075,0) .. controls (1.15,-0.25) .. (1.15,-0.5)
    arc[
        start angle=0,
        end angle=-180,
        radius=0.15
    ]
    .. controls (0.85,-0.25) .. (0.8,0.0) .. controls (0.7,0.5) .. (1,1);

\end{tikzpicture}

%% file: tikz_figs/1dim/4a.tex
\begin{tikzpicture}

    \definecolor{customred}{RGB}{219,107,67}
    \definecolor{custompurple}{RGB}{133,121,175}
    \definecolor{customyellow}{RGB}{215,159,65}


        \draw[
        teal,
        line width=0.4pt,
        dash pattern=on 0.8pt off 0.8pt
    ] (1.1,0.) arc[
        start angle=0,
        end angle=150,
        radius=0.1
    ];
    
    \draw[
        teal,
        line width=0.4pt,
        dash pattern=on 0.8pt off 0.8pt,
        decoration={
            markings,
            mark=at position 0.6 with {
                \arrow{>[scale=0.65]}
            }
        },
        postaction={decorate}
    ] (1.,-0.5) .. controls (1.1,-0.2) .. (1.1,0);

    \draw[
        teal,
        line width=0.4pt,
        dash pattern=on 0.8pt off 0.8pt,
    ] (1.,-0.5) .. controls (.9,-0.2) .. (.9,-.05);

    \draw[
        pink,
        line width=0.4pt,
        dash pattern=on 0.8pt off 0.8pt
    ] (0.9,0) arc[
        start angle=-180,
        end angle=-20,
        radius=0.1
    ];
    
    \draw[
        pink,
        line width=0.4pt,
        dash pattern=on 0.8pt off 0.8pt,
        decoration={
            markings,
            mark=at position 0.6 with {
                \arrow{>[scale=0.65]}
            }
        },
        postaction={decorate}
    ] (1.,0.5) .. controls (.9,0.2) .. (.9,0);

    \draw[
        pink,
        line width=0.4pt,
        dash pattern=on 0.8pt off 0.8pt,
    ] (1.,0.5) .. controls (1.1,0.2) .. (1.1,.05);

    \draw[
        line width=0.7pt,
        color=custompurple,
        -{To[scale=0.7]}
    ] (0,0) -- (0.8,0);

    \draw[
        line width=0.7pt,
        color=custompurple
    ] (0,0) -- (2,0);

    \node[font=\fontsize{4}{12},custompurple] at (-0.1,0.2) {$b_1$};
    \node[font=\fontsize{4}{12},custompurple] at (2.1,0.2) {$b_2$};

    \fill[custompurple] (0,0) circle (1.2pt);
    \fill[custompurple] (2,0) circle (1.2pt);

    \fill[customred] (1,-0.5) circle (1.2pt);
    \node[
        font=\fontsize{4}{12},
        customred
    ] at (0.7,-0.6) {$a_1$};


    \fill[customred] (1,0.5) circle (1.2pt);
    \node[
        font=\fontsize{4}{12},
        customred
    ] at (1.3,0.6) {$a_2$};


\end{tikzpicture}

%% file: tikz_figs/1dim/4b.tex
\begin{tikzpicture}

    \definecolor{customred}{RGB}{219,107,67}
    \definecolor{custompurple}{RGB}{133,121,175}
    \definecolor{customyellow}{RGB}{215,159,65}


    \draw[
        color=custompurple,
        line width=0.7pt,
    ] (1.1,0.5) arc[
        start angle=0,
        end angle=180,
        radius=0.1
    ];

        \draw[
        color=custompurple,
        line width=0.7pt,
    ] (1.2,0.5) arc[
        start angle=0,
        end angle=180,
        radius=0.2
    ];

        \draw[
        color=custompurple,
        line width=0.7pt,
    ] (1.1,-0.5) arc[
        start angle=0,
        end angle=-180,
        radius=0.1
    ];

        \draw[
        color=custompurple,
        line width=0.7pt,
    ] (1.2,-0.5) arc[
        start angle=0,
        end angle=-180,
        radius=0.2
    ];

    \draw[
        line width=0.7pt,
        color=custompurple,
        -{To[scale=0.7]}
    ] (0,0) -- (0.4,0);

    \draw[
        line width=0.7pt,
        color=custompurple
    ] (0,0) -- (.6,0);

    \draw[
        line width=0.7pt,
        color=custompurple,
        -{To[scale=0.7]}
    ] (1.4,0) -- (1.8,0);

    \draw[
        line width=0.7pt,
        color=custompurple
    ] (1.4,0) -- (2,0);

    \draw[
        line width=0.7pt,
        color=custompurple
    ] (0.6,0) .. controls (0.75,0) .. (.8,-0.5);

    \draw[
        line width=0.7pt,
        color=custompurple
    ] (1.2,0.5) .. controls (1.25,0) .. (1.4,0);

    \draw[
        line width=0.7pt,
        color=custompurple
        ] (1.2,-0.5) .. controls (1.2,-.25) .. (1.15,0.0) .. controls (1.1,0.25) .. (1.1,0.5);

    \draw[
        line width=0.7pt,
        color=custompurple
        ] (1.1,-0.5) .. controls (1.1,-.25) .. (1.,0.0) .. controls (0.9,0.25) .. (0.9,0.5);

    \draw[
        line width=0.7pt,
        color=custompurple
        ] (0.9,-0.5) .. controls (0.9,-.25) .. (.85,0.0) .. controls (0.8,0.25) .. (0.8,0.5);

    \node[font=\fontsize{4}{12},custompurple] at (-0.1,0.2) {$b_1$};
    \node[font=\fontsize{4}{12},custompurple] at (2.1,0.2) {$b_2$};

    \fill[custompurple] (0,0) circle (1.2pt);
    \fill[custompurple] (2,0) circle (1.2pt);

    \fill[customred] (1,-0.5) circle (1.2pt);
    \node[
        font=\fontsize{4}{12},
        customred
    ] at (0.7,-0.6) {$a_1$};


    \fill[customred] (1,0.5) circle (1.2pt);
    \node[
        font=\fontsize{4}{12},
        customred
    ] at (1.3,0.6) {$a_2$};


\end{tikzpicture}

%% file: tikz_figs/2dim/2a.tex
\begin{tikzpicture}[scale=0.8]
    \definecolor{customred}{RGB}{219,107,67}
    \definecolor{custompurple}{RGB}{133,121,175}
    \definecolor{customyellow}{RGB}{215,159,65}

    \fill[custompurple, opacity=0.18]
        (-2.8,-1.65) -- (-2.8,-0.55) -- (-0.95,-0.55) -- cycle;

    \draw[custompurple, line width=2pt] (-2.8,-1.65) -- (2.8,1.68);
    \node[custompurple] at (-2.35,-1.95) {\small\textit{$B_1$}};

    \draw[custompurple, line width=2pt] (-2.8,-0.55) -- (2.8,-0.55);
    \node[custompurple] at (-2.35,-0.25) {\small\textit{$B_2$}};

    \draw[customred, line width=2pt] (-2.6,2.4) -- (2.7,-2.1);
    \node[customred] at (2.05,-1.95) {\small\textit{$D_1$}};

    \draw[customred, line width=2pt] (-2.75,1.45) -- (2.7,2.55);
    \node[customred] at (2.2,2.85) {\small\textit{$D_2$}};

    \draw[customyellow, line width=2pt] (-1.9,2.8) -- (2.75,0.35);
    \node[customyellow] at (-1.05,2.75) {\small\textit{$T_1$}};
\end{tikzpicture}

%% file: tikz_figs/2dim/2c.tex
\begin{tikzpicture}[scale=0.8]

    \definecolor{customred}{RGB}{219,107,67}
    \definecolor{custompurple}{RGB}{133,121,175}
    \definecolor{customyellow}{RGB}{215,159,65}

    %
    \fill[
        customred,
        opacity=0.28
    ]
        (0.2826, 0.1830)
        --
        (1.1460,-0.5500)
        --
        (0.6029,-0.5500)
        --
        (-0.0368,-0.0069)
        -- cycle;

    \draw[
        custompurple,
        line width=2pt
    ]
        (-2.8,-1.65) -- (2.8,1.68);

    \node[
        custompurple
    ]
        at (-2.35,-1.95)
        {\small\textit{$B_1$}};

    \draw[
        custompurple,
        line width=2pt
    ]
        (-2.8,-0.55) -- (2.8,-0.55);

    \node[
        custompurple
    ]
        at (-2.35,-0.25)
        {\small\textit{$B_2$}};

    \draw[
        customred,
        line width=2pt
    ]
        (-2.6,2.4) -- (2.7,-2.1);

    \node[
        customred
    ]
        at (2.05,-1.95)
        {\small\textit{$D_1$}};

    \draw[
        customred,
        line width=2pt
    ]
        (-2.75,1.45) -- (2.7,2.55);

    \node[
        customred
    ]
        at (2.2,2.85)
        {\small\textit{$D_2$}};

    \draw[
        customyellow,
        line width=2pt
    ]
        (-1.9,2.8) -- (2.75,0.35);

    \node[
        customyellow
    ]
        at (-1.05,2.75)
        {\small\textit{$T_1$}};

\end{tikzpicture}

%% file: tikz_figs/2dim/2b.tex
\begin{tikzpicture}[scale=0.8]
    \definecolor{customred}{RGB}{219,107,67}
    \definecolor{custompurple}{RGB}{133,121,175}
    \definecolor{customyellow}{RGB}{215,159,65}

    \fill[customyellow, opacity=0.18]
        (-0.95,-0.55) -- (1.59062, 0.960854) -- (2.75,0.35) -- (2.75,-0.55) -- cycle;

    \draw[custompurple, line width=2pt] (-2.8,-1.65) -- (2.8,1.68);
    \node[custompurple] at (-2.35,-1.95) {\small\textit{$B_1$}};

    \draw[custompurple, line width=2pt] (-2.8,-0.55) -- (2.8,-0.55);
    \node[custompurple] at (-2.35,-0.25) {\small\textit{$B_2$}};

    \draw[customred, line width=2pt] (-2.6,2.4) -- (2.7,-2.1);
    \node[customred] at (2.05,-1.95) {\small\textit{$D_1$}};

    \draw[customred, line width=2pt] (-2.75,1.45) -- (2.7,2.55);
    \node[customred] at (2.2,2.85) {\small\textit{$D_2$}};

    \draw[customyellow, line width=2pt] (-1.9,2.8) -- (2.75,0.35);
    \node[customyellow] at (-1.05,2.75) {\small\textit{$T_1$}};
\end{tikzpicture}

%% file: tikz_figs/sunrise_diagram.tex
\begin{tikzpicture}[
    line cap=round,
    line join=round
]

\path[use as bounding box] (-5.4,-2.1) rectangle (5.4,3.0);


\definecolor{propagatorgray}{RGB}{202,202,218}


\coordinate (X2) at (-4,0);
\coordinate (X1) at ( 4,0);

\def\futureheight{2.8}


\draw[
    propagatorgray,
    line width=4pt
]
(-4.8,\futureheight) -- (4.8,\futureheight);


\draw[
    black,
    line width=1.7pt
]
(X2)
.. controls (-2.5,-1.55) and (2.5,-1.55) ..
(X1);

\draw[
    black,
    line width=1.7pt
]
(X2)
.. controls (-2.4,-0.55) and (2.4,-0.55) ..
(X1);

\draw[
    black,
    line width=1.7pt
]
(X2)
.. controls (-2.4,0.55) and (2.4,0.55) ..
(X1);

\draw[
    black,
    line width=1.7pt
]
(X2)
.. controls (-2.5,1.55) and (2.5,1.55) ..
(X1);


\draw[
    propagatorgray,
    line width=1.7pt
]
(X2) -- (-4.55,\futureheight);

\draw[
    propagatorgray,
    line width=1.7pt
]
(X2) -- (-3.45,\futureheight);

\draw[
    propagatorgray,
    line width=1.7pt
]
(X1) -- (3.45,\futureheight);

\draw[
    propagatorgray,
    line width=1.7pt
]
(X1) -- (4.55,\futureheight);


\fill[black] (X2) circle (4.8pt);
\fill[black] (X1) circle (4.8pt);


\node[
    font=\fontsize{13}{14}\selectfont,
    anchor=south east,
    xshift=-3pt,
    yshift=3pt
]
at (X2)
{$x_2$};

\node[
    font=\fontsize{13}{14}\selectfont,
    anchor=south west,
    xshift=3pt,
    yshift=3pt
]
at (X1)
{$x_1$};


\node[
    font=\fontsize{12}{13}\selectfont,
    fill=white,
    inner sep=1pt
]
at (0,-1.18)
{$y_1$};

\node[
    font=\fontsize{12}{13}\selectfont,
    fill=white,
    inner sep=1pt
]
at (0,-0.38)
{$y_2$};

\node[
    font=\fontsize{12}{13}\selectfont,
    fill=white,
    inner sep=1pt
]
at (0,0.38)
{$y_3$};

\node[
    font=\fontsize{12}{13}\selectfont,
    fill=white,
    inner sep=1pt
]
at (0,1.18)
{$y_4$};

\end{tikzpicture}

%% file: tikz_figs/acnode_diagram.tex
\begin{tikzpicture}[
    line cap=round,
    line join=round
]

\path[use as bounding box] (-5.4,-2.1) rectangle (5.4,3.0);


\definecolor{propagatorgray}{RGB}{202,202,218}


\coordinate (X3) at (-4,0);
\coordinate (X4) at (-0.45,1.35);
\coordinate (X2) at ( 0.45,-1.35);
\coordinate (X1) at ( 4,0);

\def\futureheight{2.8}


\draw[
    propagatorgray,
    line width=4pt
]
(-5.0,\futureheight) -- (5.0,\futureheight);


\draw[propagatorgray, line width=1.7pt]
    (X3) -- (-4.7,\futureheight);

\draw[propagatorgray, line width=1.7pt]
    (X3) -- (-3.7,\futureheight);

\draw[propagatorgray, line width=1.7pt]
    (X4) -- (-1,\futureheight);

\draw[propagatorgray, line width=1.7pt]
    (X4) -- (-0.4,\futureheight);

\draw[propagatorgray, line width=1.7pt]
    (X2) -- (0.5,\futureheight);

\draw[propagatorgray, line width=1.7pt]
    (X2) -- (1.35,\futureheight);

\draw[propagatorgray, line width=1.7pt]
    (X1) -- (3.7,\futureheight);

\draw[propagatorgray, line width=1.7pt]
    (X1) -- (4.7,\futureheight);


\draw[black, line width=1.7pt] (X3) -- (X4); 
\draw[black, line width=1.7pt] (X4) -- (X1); 
\draw[black, line width=1.7pt] (X3) -- (X2); 
\draw[black, line width=1.7pt] (X2) -- (X1); 
\draw[black, line width=1.7pt] (X4) -- (X2); 


\fill[black] (X1) circle (4.8pt);
\fill[black] (X2) circle (4.8pt);
\fill[black] (X3) circle (4.8pt);
\fill[black] (X4) circle (4.8pt);


\node[
    font=\fontsize{13}{14}\selectfont,
    anchor=south east,
    xshift=-3pt,
    yshift=2pt
]
at (X3)
{$x_3$};

\node[
    font=\fontsize{13}{14}\selectfont,
    anchor=south west,
    xshift=3pt,
    yshift=2pt
]
at (X1)
{$x_1$};

\node[
    font=\fontsize{13}{14}\selectfont,
    anchor=north east,
    xshift=1pt,
    yshift=-6pt
]
at (X4)
{$x_4$};

\node[
    font=\fontsize{13}{14}\selectfont,
    anchor=north,
    yshift=-5pt
]
at (X2)
{$x_2$};


\node[
    font=\fontsize{12}{13}\selectfont,
    fill=white,
    inner sep=1pt
]
at (-2.0,0.72)
{$y_3$};

\node[
    font=\fontsize{12}{13}\selectfont,
    fill=white,
    inner sep=1pt
]
at (1.75,0.72)
{$y_4$};

\node[
    font=\fontsize{12}{13}\selectfont,
    fill=white,
    inner sep=1pt
]
at (-1.85,-0.72)
{$y_2$};

\node[
    font=\fontsize{12}{13}\selectfont,
    fill=white,
    inner sep=1pt
]
at (1.80,-0.72)
{$y_1$};

\node[
    font=\fontsize{12}{13}\selectfont,
    fill=white,
    inner sep=1pt
]
at (0.02,0)
{$y_5$};

\end{tikzpicture}

%% file: tikz_figs/bubble/s=0_a.tex
\begin{tikzpicture}
    \definecolor{customred}{RGB}{219,107,67}
    \definecolor{custompurple}{RGB}{133,121,175}
    \definecolor{customyellow}{RGB}{215,159,65}



    \draw[
        thick,
        customyellow,
        decoration={
            zigzag,
            amplitude=0.5pt,
            segment length=4pt,
            pre=moveto,
            pre length=1pt
        },
        decorate
    ] (-0.5,0) -- (0,0);

    \draw[
        thick,
        customyellow,
        decoration={
            zigzag,
            amplitude=0.5pt,
            segment length=4pt,
            pre=moveto,
            pre length=1pt
        },
        decorate
    ] (2,0) -- (2.5,0);
    
    \draw[
        teal,
        line width=0.4pt,
        dash pattern=on 0.8pt off 0.8pt,
        decoration={
            markings,
            mark=at position 0.6 with {
                \arrow{>[scale=0.65]}
            }
        },
        postaction={decorate}
    ] (2,0) .. controls (1,0.2) .. (0,0);
    
    \draw[
        pink,
        line width=0.4pt,
        dash pattern=on 0.8pt off 0.8pt,
        decoration={
            markings,
            mark=at position 0.6 with {
                \arrow{>[scale=0.65]}
            }
        },
        postaction={decorate}
    ] (0,0) .. controls (1,-0.2) .. (2,0);

    \draw[
        line width=0.7pt,
        color=custompurple,
        -{To[scale=0.7]}
    ] (0,0) -- (0.8,0);

    \draw[
        line width=0.7pt,
        color=custompurple
    ] (0,0) -- (2,0);

    \node[font=\fontsize{4}{12},custompurple] at (-0.15,0.15) {$z_1^-$};
    \node[font=\fontsize{4}{12},custompurple] at (2.15,0.15) {$z_1^+$};

    \fill[custompurple] (0,0) circle (1.2pt);
    \fill[custompurple] (2,0) circle (1.2pt);

    \fill[customyellow] (2,0) circle (0.6pt);
    \fill[customyellow] (0,0) circle (0.6pt);

    \fill[customred] (1.9,-0.5) circle (1.2pt);
    \node [font=\fontsize{4}{12},customred] at (2.1,-0.6) {$-m_1^2$};

\end{tikzpicture}

%% file: tikz_figs/bubble/s=0_b.tex
\begin{tikzpicture}[/pgf/fpu/install only={reciprocal}]
    \definecolor{customred}{RGB}{219,107,67}
    \definecolor{custompurple}{RGB}{133,121,175}
    \definecolor{customyellow}{RGB}{215,159,65}



    \draw[
        thick,
        customyellow,
        decoration={
            zigzag,
            amplitude=0.5pt,
            segment length=4pt,
            pre=moveto,
            pre length=1pt
        },
        decorate
    ] (0,0) .. controls (0.2,0.2) and (1.8,0.2) .. (2.5,0);;

    \draw[
        thick,
        customyellow,
        decoration={
            zigzag,
            amplitude=0.5pt,
            segment length=4pt,
            pre=moveto,
            pre length=1pt
        },
        decorate
    ] (2,0) .. controls (1.8,-0.2) and (0.2,-0.2) .. (-0.5,0);

    \draw[
        line width=0.7pt,
        color=custompurple,
        -{To[scale=0.7]}
    ] (2,0) -- (1.2,0);

    \draw[
        line width=0.7pt,
        color=custompurple
    ] (0,0) -- (2,0);

    \node[font=\fontsize{4}{12},custompurple] at (-0.2,0.1) {$z_1^+$};
    \node[font=\fontsize{4}{12},custompurple] at (2.2,-0.1) {$z_1^-$};

    \fill[custompurple] (0,0) circle (1.2pt);
    \fill[custompurple] (2,0) circle (1.2pt);

    \fill[customyellow] (2,0) circle (0.6pt);
    \fill[customyellow] (0,0) circle (0.6pt);

    \fill[customred] (1.9,-0.5) circle (1.2pt);
    \node [font=\fontsize{4}{12},customred] at (2.,-0.6) {$-m_1^2$};

\end{tikzpicture}

%% file: tikz_figs/bubble/T_a.tex
\begin{tikzpicture}
    \definecolor{customred}{RGB}{219,107,67}
    \definecolor{custompurple}{RGB}{133,121,175}
    \definecolor{customyellow}{RGB}{215,159,65}


    \draw[
        thick,
        customyellow,
        decoration={
            zigzag,
            amplitude=0.5pt,
            segment length=4pt,
            pre=moveto,
            pre length=1pt
        },
        decorate
    ] (-0.5,0) -- (0,0);

    \draw[
        thick,
        customyellow,
        decoration={
            zigzag,
            amplitude=0.5pt,
            segment length=4pt,
            pre=moveto,
            pre length=1pt
        },
        decorate
    ] (2,0) -- (2.5,0);
    
    \draw[
        teal,
        line width=0.4pt,
        dash pattern=on 0.8pt off 0.8pt,
        decoration={
            markings,
            mark=at position 0.6 with {
                \arrow{>[scale=0.65]}
            }
        },
        postaction={decorate}
    ] (1.9,-0.5) .. controls (2.08,-0.15) .. (2.1,0);

    \draw[
        teal,
        line width=0.4pt,
        dash pattern=on 0.8pt off 0.8pt
    ] (2.1,0) arc[
        start angle=0,
        end angle=180,
        radius=0.1
    ];

    \draw[
        teal,
        line width=0.4pt,
        dash pattern=on 0.8pt off 0.8pt,
    ] (1.9,0) .. controls (1.9,-0.15) and (2,-0.35) .. (1.9,-0.5);

    \draw[
        line width=0.7pt,
        color=custompurple,
        -{To[scale=0.7]}
    ] (0,0) -- (0.8,0);

    \draw[
        line width=0.7pt,
        color=custompurple
    ] (0,0) -- (2,0);

    \node[font=\fontsize{4}{12},custompurple] at (-0.15,-0.15) {$z_1^-$};
    \node[font=\fontsize{4}{12},custompurple] at (2.2,-0.15) {$z_1^+$};

    \fill[custompurple] (0,0) circle (1.2pt);
    \fill[custompurple] (2,0) circle (1.2pt);

    \fill[customyellow] (2,0) circle (0.6pt);
    \fill[customyellow] (0,0) circle (0.6pt);

    \fill[customred] (1.9,-0.5) circle (1.2pt);
    \node [font=\fontsize{4}{12},customred] at (2.15,-0.6) {$-m_1^2$};

\end{tikzpicture}

%% file: tikz_figs/bubble/T_b.tex
\begin{tikzpicture}
    \definecolor{customred}{RGB}{219,107,67}
    \definecolor{custompurple}{RGB}{133,121,175}
    \definecolor{customyellow}{RGB}{215,159,65}


    \draw[
        thick,
        customyellow,
        decoration={
            zigzag,
            amplitude=0.5pt,
            segment length=4pt,
            pre=moveto,
            pre length=1pt
        },
        decorate
    ] (-0.5,0) -- (0,0);

    \draw[
        thick,
        customyellow,
        decoration={
            zigzag,
            amplitude=0.5pt,
            segment length=4pt,
            pre=moveto,
            pre length=1pt
        },
        decorate
    ] (2,0) -- (2.5,0);

    \draw[
        line width=0.7pt,
        color=custompurple,
        -{To[scale=0.7]}
    ] (0,0) -- (0.8,0);

    \draw[
        line width=0.7pt,
        color=custompurple
    ]
        (0,0) -- (2,0);

    \node[font=\fontsize{4}{12},custompurple] at (-0.15,-0.15) {$z_1^-$};
    \node[font=\fontsize{4}{12},custompurple] at (2.15,-0.15) {$z_1^+$};

    \fill[custompurple] (0,0) circle (1.2pt);
    \fill[custompurple] (2,0) circle (1.2pt);

    \fill[customyellow] (2,0) circle (0.6pt);
    \fill[customyellow] (0,0) circle (0.6pt);

    \fill[customred] (1.9,-0.5) circle (1.2pt);
    \node [font=\fontsize{4}{12},customred] at (2.15,-0.7) {$-m_1^2$};

        \draw[
        line width=0.7pt,
        color=custompurple,
        -{To[scale=0.7]}
    ] (1.85,-0.415) arc[
        start angle=-240,
        end angle=80,
        radius=0.1
    ];

\end{tikzpicture}